\documentclass[useAMS,usenatbib]{mn2e}
\usepackage{graphicx,epsfig,psfig,amsmath}
\usepackage{url}
\title[Chemodynamics of a Simulated Disc Galaxy]{Chemodynamics of a Simulated Disc Galaxy: Initial Mass Functions and SNIa Progenitors}
\author[Few et~al.]{C.G. Few$^{1,2}$\thanks{E-mail:c.gareth.few@gmail.com},
  S. Courty$^{3}$, B.K. Gibson$^{2,4}$,  L. Michel-Dansac$^{3}$, F. Calura$^{5}$\\
$^{1}$School of Physics, University of Exeter, Stocker Road, Exeter EX4 4QL\\
$^{2}$Jeremiah Horrocks Insitute, University of Central Lancashire, Preston, PR1~2HE, UK\\
$^{3}$Universit\'e de Lyon, Lyon, F-69003, France ; Universit\'e Lyon 1, \\ 
 Observatoire de Lyon, 9 avenue Charles Andr\'e, Saint-Genis Laval, F-69230, France \\
 CNRS, UMR 5574, Centre de Recherche Astrophysique de Lyon; Ecole Normale Sup\'erieure de Lyon, Lyon, F-69007, France \\
$^{4}$Department of Astronomy \& Physics, Saint Mary’s University, Halifax, NS B3H 3C3, Canada\\
$^{5}$INAF, Osservatorio Astronomico di Bologna, via Ranzani 1, 40127, Bologna, Italy\\
}

\begin{document}
\date{Accepted}
\pagerange{\pageref{firstpage}--\pageref{lastpage}} \pubyear{2014}
\maketitle
\label{firstpage}

\begin{abstract}
We trace the formation and advection of several elements within a cosmological adaptive mesh refinement
simulation of an L$\star$ galaxy. We use nine realisations of the same
initial conditions with different stellar Initial Mass Functions
(IMFs), mass limits for type-II and type-Ia supernovae (SNII, SNIa)
and stellar lifetimes to constrain these sub-grid phenomena. Our code
includes self-gravity, hydrodynamics, star formation, radiative
cooling and feedback from multiple sources within a cosmological
framework. Under our assumptions of nucleosynthesis we find that SNII with progenitor masses of up to 100 M$_\odot$
are required to match low metallicity gas oxygen abundances. Tardy
SNIa are necessary to reproduce the classical chemical evolution
‘knee’ in [O/Fe]-[Fe/H]: more prompt SNIa delayed time distributions
do not reproduce this feature. Within our framework of hydrodynamical
mixing of metals and galaxy mergers we find that chemical evolution is
sensitive to the shape of the IMF and that there exists a degeneracy
with the mass range of SNII. We look at the abundance plane and
present the properties of different regions of the plot, noting the
distinct chemical properties of satellites and a series of nested discs that have greater velocity dispersions, are more $\alpha$-rich and metal poor with age.

\end{abstract}


\section{Introduction}

The abundance of different elements is a key astrophysical observable,
with variations occuring over a large range of spatial and temporal
scales which provide deep insights into the formation and evolution of
galaxies. Modelling the chemical evolution of galaxies is an extremely
useful tool largely because the differing nucleosynthetic origins of chemical
species created in stars tells us a great deal about the large scale
assembly not only of galaxies but also of clusters.

Complex theoretical models of nucleosythesis \citep{iben78, arnett78,
  chiosi79,nomoto84, maeder92,ww95,vdhoek97,portinari98,iwamoto99,marigo01,limongi03,izzard04,chieffi04,kobayashi06,
  karakas07,karakas10, doherty10} tell us how the mass of
different stars influences the quantity of different elements that are
synthesised during their lifetime and any subsequent supernova (SN)
nucleosynthesis that takes place. Through various ejection
mechanisms, such as SN mass ejection or stellar winds, these elements
are recycled into the interstellar medium (ISM) from which subsequent 
generations of stars may form. The characteristic abundance ratios of
nucleosynthetic sources with different lifetimes results in an
evolving chemical signature in the ISM which is imprinted onto stars
and may be observed, particularly in long-lived stellar populations \citep{carbon87,
edvardsson93, bensby05, reddy06}.
Further complications to this framework come in the form of
dilution by gas inflow, loss of metal-rich gas through galactic
outflows/fountains, galaxy mergers and the internal motion of gas and
stars, all of which are rich in detail and complex to model.

Chemical evolution models (CEMs) began as `semi-numerical' approaches with
parametrisations of star formation, stellar lifetimes, nucleosynthesis
and gas dilution \citep[e.g.][]{talbot71,pagel75,tinsley80,matteucci89,carigi94,gibson97,chiappini97,ramirez07,calura10}.
Previous attempts to add a cosmological context to chemical evolution modelling used semi-analytical models 
set in a Lambda-CDM background \citep{calura09, yates13}. Both semi-numerical and semi-analytical approaches 
are powerful tools for constraining the aspects of input physics considered, but the limitations of these tools are that 
they do not follow the hydrodynamics of the gas, the advection of metals, or easily track physical processes on sub-galactic scales. We will show later that features exist in the abundance ratio plane 
that do not arise without hydrodynamical simulation.
The ever increasing capacity of high performance computers to run elaborate
simulations has allowed CEMs to be combined with dynamical simulations 
\citep{lia02,valdarnini03,kawata03,kobayashi04,tornatore04,romeo05,martinezserrano08a,oppenheimer08,wiersma09,shen10}
in what is referred to as `chemodynamics'.
All these models are based upon a Lagrangian approach
(smoothed particle hydrodynamics) with little or no attention given to
Eulerian approaches. In \cite{few12} we sought to remedy this,
with the presentation of modifications to the Adaptive Mesh Refinement
(AMR) code \textsc{ramses} that allowed the chemodynamics to be
simulated in an approach that complements existing Lagrangian codes.

The significance of chemodynamics is that while the way that 
the initial mass function (IMF) and star formation history give rise to
chemical evolution is clear, it can also be demonstrated that the chemical
evolution indirectly influences the hydrodynamics of the ISM. The
radiative cooling rates of plasmas depends upon the chemical
composition and has a knock-on effect on metal transport in galaxies
\citep{scannapieco05}. This non-linear feedback between element
production and star formation is linked to the well known energetic
feedback loop between supernovae and star formation in dynamical simulations.

In this work we consider the influence of the IMF and SNIa progenitors. The IMF defines the fraction 
of stars in each mass interval, controlling the weighting of each nucleosynthetic 
source and the strength of different energy sources in SN feedback. A qualitative 
description of the IMF was first formalised by \cite{salpeter55} as a description 
of the “original mass function” but numerous multi-slope IMFs \citep{tinsley80, scalo86,
 kroupa93, scalo98, kroupa01} have now been proposed and present a 
predicted luminosity function that is much closer to observations. At the present time, separate studies 
are converging towards a stellar IMF with a log-normal form proposed by \cite{chabrier03} which is
similar to the multi-power law IMF of \cite{kroupa01} used in this study.

Our understanding of an important site of nucleosynthesis, SNIa
progenitors, is lacking with regard to the lifetime of the progenitors.
In this work we do not make any specific assumption concerning the
nature of the progenitors by using a parametrisation 
of the time-scale of SNIa called the delayed time distribution (DTD) which may be determined empirically. This 
DTD is the normalised SNIa rate of a simple stellar population (SSP).

The metal enrichment in numerical simulations has been followed in a
similar manner to the work presented here for the IGM/ICM
\citep{tornatore07, wiersma09, tornatore10}, elliptical galaxies
\citep{kawata03} and disc galaxies \citep{kobayashi11, brook12, bekki13, martel13, agertz13}. Most of
these works employ Lagrangian methods for the numerical hydrodynamics
and to date few studies of the influence of the IMF and SNIa formalism
has been conducted with grid codes. Conventional treatments of
SPH suffer from well known issues in resolving hydrodynamical
instabilities in certain regimes \citep{agertz07,tasker08} and also lack implicit
diffusion of metals between local particles. It must be said that great
success has been had with the inclusion of turbulent mixing models
(Shen et al. 2010), yet the diffusion coefficient associated with this mixing has a
strong influence on (for example) the galactic metallicity
distribution function \citep{pilkington12c}. AMR codes not only provide
an essentially natural diffusion via mixing but are also very capable in
resolving instabilities \citep{agertz07,tasker08}. Other code
comparison papers \citep{frenk99,oshea05,house11,pilkington12a,scannapieco12} 
note the variable results that come from galaxies evolved with different codes. 

In this work we use the \textsc{ramses-ch} code presented in \cite{few12} to compare
multiple realisations of a single galaxy in a cosmological environment
that use different sub-grid CEMs with the aim of providing constraints
on the IMF and SNIa progenitors. We will also demonstrate that the
cosmological assembly of a galaxy leaves important signatures within
its chemical abundance properties. We vary the IMF over a range of
slopes that covers those used in the literature and also change the SNIa
scheme to compare three models with different delayed time
distributions. Additional models are used to examine the influence of
the SNII mass range and the binary fraction.

The layout of the paper is as follows: In \S\ref{method} we describe the initial conditions, 
numerical method and chemical evolution models employed in this work. In \S\ref{results} we 
demonstrate a kinematic decomposition of the galaxy disc and spheroid, detail the general structural 
properties of the realisations and present the star formation and SN rate histories, abundance 
space diagrams and decompose the abundance plane into different formation episodes. \S\ref{discuss} is a 
discussion of the results followed by our conclusions in \S\ref{conc}.

\section{Method}
\label{method}

\subsection{Code}

In this work we use \textsc{ramses-ch} \citep{few12}
a chemodynamical version of \textsc{ramses} \citep{teyssier02} to simulate a disc galaxy in a
cosmological context with subgrid physics accounting for star
formation, supernovae and chemical evolution.\footnote{We are using version 3.07 of \textsc{ramses}.} Dark matter and stars
move in the gravitational potential as N-body particles while the
hydrodynamical evolution is followed by an adaptive grid that
automatically increases the resolution for local overdensities.
The hydrodynamics grid is used to advect the abundance of elements 
H, C, N, O, Ne, Mg, Si, Fe and the global metallicity Z.

Star particles are permitted to form in gas with a density $n_\mathrm{g}
>n_{0}$ ($=0.3$~cm$^{-3}$) and are spawned stochastically \citep{dubois08} 
at an average rate of $\dot\rho_* =\epsilon_*\rho_\mathrm{g}/t_\mathrm{ff}$, 
where $t_\mathrm{ff}=(3\pi/32G\rho_\mathrm{g})^{1/2}$ is the local free-fall time
of gas, $\rho_\mathrm{g}$ is the gas mass density and $\epsilon_*$ (=0.01) is the
star formation efficiency which is constrained such that the
simulation can reproduce the Schmidt-Kennicutt relation \citep{schmidt59,kennicutt98}.

In each timestep star particles eject mass and enrich the surrounding gas as
well as injecting energy from SNe. Each star particle has a mass of 
8$\times$10$^{5}$~M$_\odot$ and represents a simple stellar population
(SSP). During the SNII phase ($\sim$30~Myr) the most
massive stars explode and energy is added to the gas within a sphere
that is two grid cells in radius. This energy is added kinetically
with a blastwave profile (the velocity of the blast wave is linearly
interpolated with radius) that has a total energy of $E_\mathrm{g,K} =
\epsilon_\mathrm{SN} E_\mathrm{SN} N_\mathrm{SNII}$ where $N_\mathrm{SNII}$ is the number of SNII
in the star particle exploding in a given timestep and $\epsilon_\mathrm{SN}$ is the efficiency
with which the energy per SN ($E_\mathrm{SN}=10^{51}$~erg) couples to the
surrounding ISM. An efficiency of $\epsilon_\mathrm{SN}=1$ is used
throughout this work. We find that using $\epsilon_\mathrm{SN}=0.5$
leads to an increased stellar mass fraction (of the total mass in the
virial radius) from 0.08 to 0.09. As we show later this is not a
large change compared to that which arises through changing our
CEM but it is somewhat dependent on the other input parameters so
greater stellar mass should not be unexpected if choosing a lower value of
$\epsilon_\mathrm{SN}$ than we use here.
The super-bubble size of two grid cells corresponds to a physical 
size of 872~pc which is large compared to the true size of super-bubbles of $\sim$100~pc but 
necessary to implement effective kinetic feedback.
The velocity of the blastwave depends upon the mass of the ejecta
which may include matter that is swept up, in addition to the mass ejected
from stars. This swept up matter is parametrised by $f_\mathrm{w}$ 
such that the total mass of gas outflowing from a star particle is $(1+f_\mathrm{w})m_\mathrm{ej}$ where $m_\mathrm{ej}$ is the
mass ejected by the stars themselves. Mass-loading the ejecta with this swept up material couples the ISM to feedback 
more strongly and is necessary to produce galactic outflows. The
galactic outflows are necessary for there to be any enrichment of the
IGM in our model, however the downside is that the velocity of the
ejected material is diminished as the mass increases such that as the
mass-loading factor becomes high enough the feedback becomes unable to
effectively remove mass from star formation regions. Having tested
several values of $f_\mathrm{w}$ ranging from 0 to 20 we adopted $f_\mathrm{w}$=10 for all
models as a compromise that allows IGM enrichment without unduly
affecting the ability of feedback to reduce the stellar mass of the
galaxy. No more than 25\% of a gas
cell's mass can be removed by mass loading in order to prevent excessively depleting a gas cell.

Later in the star particle's life it ceases to produce SNII ejections and instead
produces SNIa ejections and also AGB stellar winds. We will show in
\S\ref{chemevo} that the number of SNIa is much lower than that of
SNII and for this reason we do not employ the same mass/energy
injection method for SNIa. This later phase of the star particle's
existence is also shared by AGB stellar winds which enrich the local
gas. We combine the two sources of mass into a combined AGB/SNIa
feedback source, depositing mass and energy (in a thermal form) to the
local gas cell only. The energy injected in this situation is simply $E_\mathrm{g,T} =
\epsilon_\mathrm{SN} E_\mathrm{SN} N_\mathrm{SNIa}$. We plan to explore the 
possibility of implementing kinetic feedback for SNIa in future work.
While running tests of the code we ran one of our simulations with no
SNIa feedback, in this case we found that the stellar mass forming in
the halo was slightly higher, but the increase was no larger than the
amount we find for differences between our models which are discussed later.

For gas with a high density and low temperature there is a danger that
the Jeans length of the gas will be poorly resolved and introduce
numerical (and undesirable) fragmentation to the gas. It can be
impractical to refine the gas grid sufficiently to provide such
resolution in every situation for these simulations. To prevent
unphysical fragmentation of the gas we introduce a polytropic equation
of state for cool gas ($T<T_\mathrm{th}$) in the
form $T = T_\mathrm{th}(n_\mathrm{g}/n_{0})^{\gamma-1}$. At the resolution employed in
this work an index of $\gamma=2$ and a temperature threshold $T_\mathrm{th}=2900$~K allows the Jeans 
length to be resolved by at least four grid cells at all times as advised in \cite{truelove97}.

Radiative cooling of the gas is computed assuming photo-ionisation
equilibrium as a function of temperature for different metallicities
and densities. A uniform UV background \citep{haardt96} that has a 
dependence on redshift provides a heat source. The redshift dependent 
contribution to cooling from metals in a photoionised gas is calculated, 
for temperatures above 10$^4$~K, via a fit of the difference between the 
cooling rates at solar metallicity and those at zero metallicity using 
photoionisation code cloudy \citep{ferland98}. For gas cooler than 10$^4$~K, 
metal fine-structure cooling rates are taken from \cite{rosen95}.
Note that the polytropic equation of state that is used in these simulations
prevents gas from remaining in this low temperature regime.

\subsection{Chemical evolution}
\label{chemevo}

The subgrid chemical evolution model that \textsc{ramses-ch} uses
incorporates the IMF, nucleosynthesis models and progenitor lifetimes
for each of SNII, SNIa and AGB. These are used to
calculate (for different initial metallicities) what the cumulative
(as a function of stellar population age) number of SNII, SNIa and AGB
is. The table is created in a cumulative form to allow the simulation
timestep (which is variable) to be completely
independent of the feedback table age-step; by doing so we need only
know the number of, for example, SNII that have exploded in it
so far. Using the star particle's age we can look up the number of SNII
that should have exploded over its entire lifetime and take the
difference to determine how many SNII are due in the current
timestep. The same method is followed to determine the mass of each
element considered and the total ejected mass for each of the
nucleosynthetic sources.

\subsubsection{Nucleosynthesis yields}
We use the nucleosynthetic yields for SNII in \cite{ww95} where we
adopt the yields associated with the Model B explosion energies for
stars with mass greater than 30~M$_\odot$. We have also applied a
correction suggested by \cite*{timmes95} by halving the \cite{ww95} Fe yield.
Yields for AGB stars are taken from \cite{vdhoek97}. This pair of models is a common choice and when
employed in semi-numerical models by \cite{romano10} were found to reproduce observed Milky Way values of [O/Fe] 
and [C/Fe] reasonably well, although the same work found that a
different combination of yields \citep{kobayashi06,karakas10} performed better overall when
considering the majority of elements. Another comparison of nucleosynthesis models was conducted
by \cite{francois04} where \cite{ww95} yields were found to give the best fits
to observations when compared with yields of solar abundance stars from \cite{nomoto97} and \cite{limongi03}; the yields for some elements required modification but in this work we focus upon O and Fe, 
both of which were found to fit observations well. Stars in the mass range 0.1--8~M$_\odot$ contributed
elements in the AGB wind channel and stars from 8~M$_\odot$ up to
$m_\mathrm{SNII,u}$ (the upper mass limit on SNII progenitors) are
taken as SNII, this upper mass limit is one of the variables in this work. Stars that are more massive than $m_\mathrm{SNII,u}$
are assumed to form but collapse into remnants in the form of black
holes and remain in the collisionless stellar particle. Details
concerning our input nucleosynthesis can be found in Appendix~A.

\subsubsection{Initial mass functions}

\begin{figure}
\includegraphics[width=84mm]{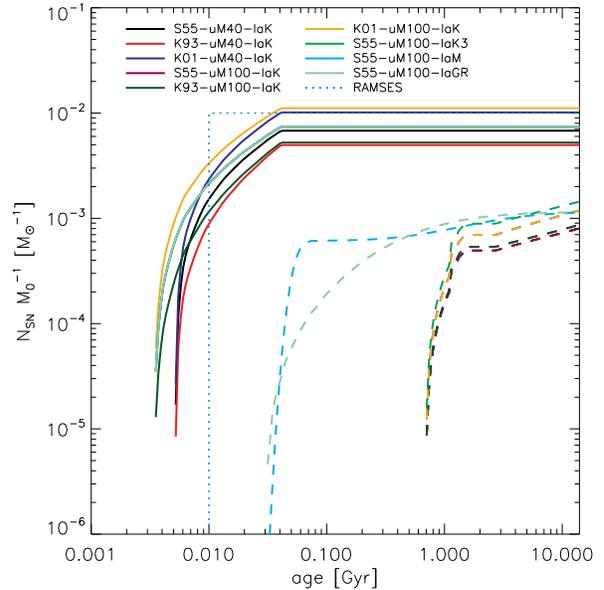}
\caption{Cumulative SN rates per solar mass formed for a stellar
  particle (with solar metallicity) for the models tested. Solid lines
  are the rates of SNII and dashed lines are SNIa, the light blue
  dotted line is the equivalent SN rate for a standard \textsc{ramses}
  run that parametrises an S55 IMF i.e. $\eta_{SN}$=0.1. The SNII lines of models with the same mass limits ($m_\mathrm{SNII,l}$ and $m_\mathrm{SNII,u}$) and IMF
  overlay one another and are not visible, i.e. models with a name
  starting “K01-uM40” have the same SNII rates.}
\label{fig:feedback}
\end{figure}

The IMF defines the number of stars that form in a particular mass
interval and its shape determines the relative dominance of each nucleosynthetic
source. In this work we have selected three IMFs for comparison, which
represent the range of slopes that have been proposed by the
community. These models are (in ascending order of bias toward stars
with masses greater than 8~M$_\odot$), Kroupa et al. (1993),
\cite{salpeter55} and \cite{kroupa01}.\footnote{This IMF is that 
from Kroupa 2001, p.~234 and not the revised version given later in the paper.}
Henceforth we refer to the \cite{kroupa01} IMF as being `top-heavy'
since it is the most biased to massive stars. We have also varied the upper
mass limit of SNII progenitors (this does not reflect a change in the
IMF merely the lifting of a restriction on progenitor masses) for some
models.

The number of SNII in a star particle of initial mass $M_\mathrm{0}$ is
calculated by integration over the IMF by number $\phi(m)$ as follows; 
\begin{equation}
  N_\mathrm{SNII}(\tau_*,Z_0) =
  M_0\int_{MAX(m_\mathrm{SNII,l},m_{\mathrm{TO}})}^{m_{\mathrm{SNII,u}}} \phi(m) {\mathrm{d}}m, 
\end{equation}
where $m_{\mathrm{TO}}$ is the main sequence turn-off mass depending on the
main sequence lifetime $\tau_*$ and initial metallicity $Z_0$. This
equation is clearly only applicable while $m_{\mathrm{TO}}(\tau_*,Z_0)$ is
in the sensible range of values for SNII progenitors,
$m_{\mathrm{SNII,l}}$ (=8~M$_\odot$) to $m_{\mathrm{SNII,u}}$. The cumulative SNII rate given by 
this equation is shown in Fig.~\ref{fig:feedback}.

AGB stars are assumed to expel their outer layers in a short period of
time (during thermal pulsation) compared with the simulation timestep and we calculate the number
of AGB stars ejecting their mass in the same way we do for the SNII:
\begin{equation}
  N_\mathrm{AGB}(\tau_*,Z_0) = M_0\int_{MAX(m_\mathrm{AGB,l},m_{\mathrm{TO}})}^{m_{\mathrm{AGB,u}}} \phi(m) {\mathrm{d}}m, 
\end{equation}
while $m_{\mathrm{TO}}$ is in the range of AGB masses,
$m_{\mathrm{AGB,l}}$ to $m_{\mathrm{AGB,u}}$ which are 0.5~M$_\odot$
and 8~M$_\odot$ respectively. To determine turn-off mass as a function of stellar population age we
use stellar lifetime models by \cite{kodama97} in which the stellar
lifetime is a function of both mass and $Z_0$.

\subsubsection{Type Ia supernovae progenitors}
A different treatment is applied to SNIa; the abundance of elements in the
SNIa ejecta is taken as a constant for which we use the yields from
\citet{iwamoto99} (shown in the right-hand panel of Fig.~\ref{fig:ejecta}). 
Three SNIa progenitor models are considered in
this work with varying time-scales. The first is taken after
\cite{greggio83} and models C/O WD progenitors, giving them an
onset time of $\sim$30~Myr. In the simulation descriptors of this work,
this model is denoted with \emph{IaGR}.

The second SNIa model considered here is by \cite*{mannucci06} and has
a more general form with a prompt and a tardy component and is denoted
by \emph{IaM} in simulation titles. The \emph{IaGR} and \emph{IaM} models 
define a time distribution for SNIa but do not explicitly give the number 
of SNIa until scaled by the number of binary systems that give rise to SNIa explosions.
This number is a relatively free parameter but can be chosen such that the present day SNIa 
rate is reproduced. We have chosen to constrain 
the number of SNIa to a value that reproduces the SNIa rates measured
in the Milky Way when the DTD is applied in a semi-numerical scheme 
(Chiappini et al. 1997; Portinari et al. 1998).

The third model is based on \cite*{hachisu99} where the mass range of
the SNIa progenitor secondaries is suggested to be bimodal (and is similar to that
used in \cite*{kobayashi00} and \citealt{kawata03}). In this model, SNIa
systems are binaries with a primary mass range of
$m_{P,l}=3$~M$_\odot$ to $m_{P,u}=8$~M$_\odot$ which evolves into a
C/O WD. Secondaries are either `main sequence' (MS) or red giant (RG)
with mass ranges of $m_{MS,l}=1.8$~M$_\odot$ to $m_{MS,u}=2.6$~M$_\odot$
and $m_{RG,l}=0.9$~M$_\odot$ and $m_{RG,u}=1.5$~M$_\odot$ respectively.
The binary fraction for each of the secondary types is $b_{MS}=0.05$
and $b_{RG}=0.02$ after \cite{kawata03}. This two component SNIa model is similar to the \cite{mannucci06}
model except that the typical
mass (and hence time-scale) of the two components are very different;
this model has an onset time of 700~Myr. The rate of
SNIa in this case is found by a double integration of the IMF and an assumed
binary fraction for the two mass ranges, 

\begin{align}
\label{eq:IaK}
  N_\mathrm{SNIa}(\tau_*) &=          M_0\int_{m_{\mathrm{P,u}}}^{m_{\mathrm{P,l}}} \phi(m) {\mathrm{d}}m \nonumber\\
                                      &\times \Biggr[b_\mathrm{MS} \frac{\int_{MAX(m_{\mathrm{MS,l}},m_{\mathrm{TO}})}^{m_{\mathrm{MS,u}}}\phi(m) {\mathrm{d}}m}{\int_{m_{\mathrm{MS,l}}}^{m_{\mathrm{MS,u}}} \phi(m) {\mathrm{d}}m} \\
                                      &+         b_\mathrm{RG}
                                      \frac{\int_{MAX(m_{\mathrm{RG,l}},m_{\mathrm{TO}})}^{m_{\mathrm{RG,u}}}
                                        \phi(m)
                                        {\mathrm{d}}m}{\int_{m_{\mathrm{RG,l}}}^{m_{\mathrm{RG,u}}}
                                        \phi(m)
                                        {\mathrm{d}}m}   \Biggr]. \nonumber
\end{align}

This SNIa model is henceforth denoted by \emph{IaK}; however, note that
unlike in the works that inspire this model
\citep{kobayashi00,kawata03} the IMF for the secondary
stars is the same as for the primaries and we have also not applied
the metallicity floor imposed on these objects ([Fe/H]$\geq$-1.1) as suggested by \cite{kobayashi00}. 
The cumulative SNIa rate for a star particle (that has an instantaneous burst of
formation) for these three models are shown in
Fig.~\ref{fig:feedback} which illustrates the differing time-scales
of each model.

To construct the chemical evolution models used in this work we select
an IMF, an upper mass range and a Ia formalism. The naming convention
for these combined models starts with either \emph{S55} \citep{salpeter55}, \emph{K93}
(Kroupa et al. 1993) or \emph{K01} \citep{kroupa01} to denote the IMF, the
second part of the name gives the upper mass limit for SNII in M$_\odot$ preceeded
by `\emph{uM}'. The last part of the name gives the Ia formalism chosen
where \emph{IaGR} is \cite{greggio83}, \emph{IaM} is \cite{mannucci06}
and \emph{IaK} is described in Eq.~\ref{eq:IaK}. We have run models
with two different binary fractions for the \emph{IaK} model where one
is enhanced to have more SNIa by a factor of 1.8. Table~\ref{tab:models} gives an 
account of the model components and the binary fractions used in the
\emph{IaK} models. The chosen binary fractions are well within the limits compiled 
by \cite{maoz08} for the number of stars with masses 3--8~M$_\odot$ that
explode as SNIa of 2-40\%. Our model \emph{IaK} has 4.7\% and \emph{IaK3} has 8.46\%, 
a value similar to the 7\% employed in the chemodynamical simulations
of \cite{kobayashi11}. We show two additional models in Appendix~B 
which both employ this enhanced \emph{IaK3} model with a \emph{K01} IMF, these are provided
only to demonstrate the effect on another IMF in addition to that used in
the main text.

\begin{table}
\begin{center}
\caption[Chemical evolution models tested in \textsc{ramses-ch}]
{Chemical evolution models used in the galaxy realisations, see the main text for details. Column (1): Galaxy ID;
  column (2): initial mass function; column (3): upper mass limit of SNII progenitors; column (4): SNIa progenitor model; 
column (5-6): binary fractions used in Eq.~\ref{eq:IaK} for the \emph{IaK} SNIa model.
}\label{tab:models}  
\begin{tabular}{l c c c c c}
Realisation name    & IMF                & $m_\mathrm{SNII,u}$ & SNIa DTD & $b_\mathrm{RG}$ & $b_\mathrm{MS}$       \\
                       &                       & (M$_\odot$)                  &                 &                          &                                 \\
\hline
S55-uM40-IaK       & S55  & 40                    & {IaK}   & 0.02 & 0.05 \\
S55-uM100-IaK     & S55  & 100                    &{IaK}  & 0.02 & 0.05\\
K93-uM40-IaK   & K93    & 40                    & {IaK} & 0.02 & 0.05\\
K93-uM100-IaK & K93    & 100                  & {IaK} & 0.02 & 0.05\\
K01-uM40-IaK   & K01    & 40                    & {IaK} & 0.02 & 0.05\\
K01-uM100-IaK & K01    & 100                  & {IaK} & 0.02 & 0.05\\
S55-uM100-IaK3 & S55    & 100                    & {IaK}   & 0.036 & 0.09     \\
S55-uM100-IaM  & S55    & 100                    & {IaM}  & -- & --    \\
S55-uM100-IaGR & S55    & 100                   & {IaGR} & -- & -- \\
\end{tabular}
\end{center}
\end{table}

\subsection{Simulation framework}

The nine chemodynamical simulations that are presented here differ with respect to the CEM they employ but all have the same initial conditions as used in \cite{sanchezblazquez09} and in \cite{few12} to simulate \emph{K01-uM40-IaK} which is included in this work. The galaxy is taken from a cosmological volume
20~$h^{-1}$~Mpc in size with cosmological parameters as follows:
$H_0$=70 km~s$^{-1}$~Mpc$^{-1}$, $\Omega_m$=0.3, $\Omega_\Lambda$=0.7, $\Omega_b$=0.042, and $\sigma_8$=0.92.
An initial cosmological (with dark matter only) run was performed and
a halo with a relatively quiet merger history was selected. This halo
was then the subject of a multi-resolved simulation with baryons included. 
Dark matter particles have a mass of 6$\times10^6$~M$_\odot$ in the central region (mass resolution
declines further from the box center to reduce computing time), in comparison a typical stellar particle 
has a mass of $\sim$8$\times10^5$~M$_\odot$. The physical 
spatial resolution is 436~pc at grid level=16. The gas is initiated with a hydrogen content of 75\%.

The virial mass of the galaxy halo is approximately
5.6$\times$10$^{11}$~M$_\odot$ and while it exists in a relatively
isolated environment it does experience several mergers from small
satellite dwarfs.

\section{Results}
\label{results}

\subsection{Kinematic decomposition}

To make a fair comparison with observed disc stars we perform a
kinematic decomposition to separate halo and bulge stars from the 
disc star population using the method employed in \cite{abadi03}, \cite{few12b} and \cite{calura12}.
While this method does not completely exclude 
bulge/halo stars from the disc population, it
does significantly reduce contamination by removing those that have 
kinematic properties clearly inconsistent with the disc. 

For each star in the sample the orbital circularity is calculated,
i.e. the ratio $J_z$/$J_{circ}$ where $J_z$ is the angular momentum
and $J_{circ}$ is the circular orbit angular momentum for a given
particle energy. The distribution is assumed to have 
a peak at $J_z$/$J_{circ}=0$ and at $J_z$/$J_{circ}=1$ which
correspond to the spheroidal and disc populations respectively.
The spheroidal component of the distribution 
is assumed to be symmetrical about $J_z$/$J_{circ}=0$, particles with
negative values of $J_z$/$J_{circ}$ are designated as spheroid stars. Particles  
with positive $J_z$/$J_{circ}$ are attributed to the 
spheroidal component with a probability proportional to the assumed
distribution of spheroid stars. Thus stars are more likely
to be designated as part of the spheroid if they are closer to the $J_z$/$J_{circ}=0$ peak. All
remaining stars are designated `disc stars'. The disc fraction of 
each realisation is given in Table~\ref{tab:results}: all are in
the range 0.81--0.86.

\subsection{Structural properties}

\begin{figure*}
\includegraphics[width=1.0\textwidth]{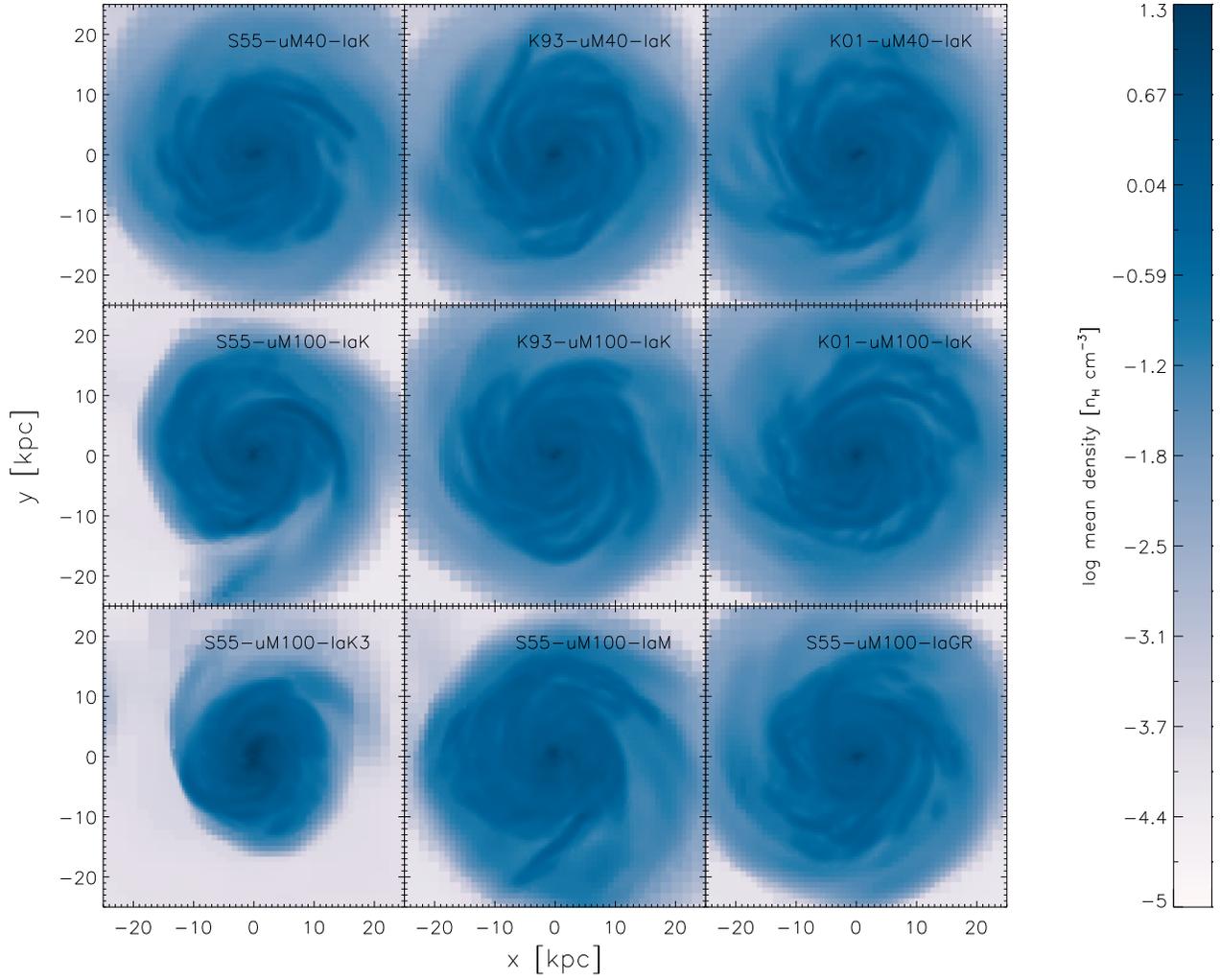}
\caption{Gas density maps of the simulated galaxies at z=0. Each pixel is coloured by the mean line-of-sight 
density.}
\label{fig:maps}
\end{figure*}

The galaxy studied here is a field environment disc galaxy with a virial mass
of approximately 5.6$\times$10$^{11}$~M$_\odot$. It experiences several
minor mergers and has a small satellite nearby at $z$=0. The satellite is positioned at 
a radius of 60~kpc and has a total mass of 7.3$\times$10$^{9}$~M$_\odot$.

In Fig.~\ref{fig:maps} we plot gas density maps (where each pixel
represents the mean line-of-sight density) for each of the
realisations to show how the sub-grid physics affects the shape 
of the spiral structure. Despite changes to the sub-grid model, only
slight differences are apparent for most of the realisations. The
obvious exception to this is \emph{S55-uM100-IaK3}. This does not stem
from changes in the merger history but rather from the additional
thermal heating applied to the gas via SNIa feedback which reduces the
reservoir of cold gas from which stars may form in the outer regions of the
disc. For a comparison of the relative strength of SNIa feedback of
the models see Fig.~\ref{fig:feedback}. \emph{S55-uM100-IaK3} is smaller than the other discs throughout its history and has the
lowest gas fraction of all the realisations, and in keeping with
having a smaller disc also has a shorter stellar scale length.

We measure the stellar disc exponential scale length for the galaxies across a radial
range between 5 and 15~kpc. This outer edge is in most cases the radius of the break
in their type II density profile (for which an outer disc with a
shorter scale length exists), i.e. we measure the scale
length of the inner disc and not the outer disc which has a steeper profile.
Scale lengths are found to have an average of 3.39~kpc, values for each galaxy can be found in
Table~\ref{tab:results}.

The galaxies have mass-weighted stellar [Fe/H] gradients of $\sim$-0.01~dex~kpc$^{-1}$, 
consistent with observed gradients
in spiral galaxies \citep*{zaritsky94}. We note that despite these
gradients having realistic values they are rather flat
which may be symptomatic of the feedback scheme distributing SNII
elements over a sphere of radius 872~pc. Another cause of
flat metallicity gradients was shown in \cite{pilkington12a} where the
star formation was shown to be quite uniform with radius which lead
very naturally to flat metallicity gradients even before stellar
migration. This uniformity of star formation is likely also due to
resolution and hence smaller feedback
spheres (available at greater resolution) may result in steeper
metallicity gradients. The stronger feedback employed in other simulations 
has been shown to lead to flatter metallicity gradients \citep{gibson13}, particularly 
at high-redshift. Constraints on high-redshift metallicity gradients
\citep{maciel03,yuan11, jones13} will reveal whether this is desirable
or whether the powerful feedback required by simulations to meet 
observed properties destroy the high-redshift gradients that we wish to reproduce.
 
The baryon fraction within the virial radius of the galaxies is not
significantly altered by any aspect of the feedback as considered in
this work, remaining at $\sim$15\% for most of the realisations. 
The general robustness of the baryon fraction under the
changes to the CEM studied here indicates that the infall rate of 
gas into the halo for a galaxy of this mass is unaffected by the changes we make to our 
numerical feedback scheme, which only redistributes baryons within the halo.

While the baryon fraction of the galaxy remains relatively unaffected by changes
in the IMF, we do see some small changes in the stellar fraction. The star formation efficiency of 
the galaxy as a whole is affected by the IMF since different IMFs imply different fractions of 
gas returned to the ISM; by replenishing the ISM more fuel is
available for star formation.
The \emph{S55} runs have the most efficient conversion of gas into stars,
with stellar mass fraction of \emph{S55-uM100-IaK} at 0.088, this is followed by
\emph{K93-uM100-IaK} runs at 0.084, \emph{K01-uM100-IaK} has the
lowest stellar fraction at 0.078. Bearing in mind that the \emph{S55}
IMF gives rise to a greater number of SNII than does a \emph{K93} IMF
this means that increasing the efficiency of the feedback channels in a model does not necessarily result in 
a reduction in the star formation rate either because it does couple
to the ISM in the right way or because the gas 
available for star formation is replenished by feedback. \cite{guo10} calculates the expected stellar 
mass as a function of halo mass from which we calculate the target stellar mass fraction 
for the dark matter halo simulated in this work as being 0.035. As expected, the stellar 
mass is too high, but only by a factor of 2.5.

The disc fraction is measured by taking the fraction of stellar mass that is 
attributed to the disc by our kinematic decomposition. All the galaxies are disc 
dominated with the disc accounting for more than 80\% of the total stellar mass.
The disc fractions of \emph{uM100} runs are consistently higher than those of \emph{uM40} 
runs by around 1\%, this is probably due to the increased level of
feedback that redistributes low angular momentum material as suggested
by \cite{binney01} and demonstrated in \cite{brook11, brook12c}.

\begin{table*}
\begin{center}
\caption[alternative caption]
{Some properties of the galaxies taken at z=0. Column (1): Galaxy ID;
  column (2): stellar disc scale length; column (3): metallicity gradient of stellar disc;
  columns (4-5): mass fractions calculated within the virial radius;
  column (6-7): SNII and SNIa rates averaged over the last 3~Gyr for
  stars within the virial radius, SNuM is the number of SN per 10$^{10}$~M$_\odot$ stellar mass; column (8): stellar disc mass fraction (of
  total stellar mass).
}\label{tab:results}  
\begin{tabular}{l c c c c c c c c}
Realisation name & scale length & d[Fe/H]/dr & stellar & gas  & SNII rate & SNIa rate & stellar disc  \\
                            & [kpc]                    &      [dex/kpc]   & fraction & fraction      & [SNuM]   & [SNuM] &  fraction \\
\hline
S55-uM40-IaK    & 3.37            &  -0.011     &0.086           &  0.063       & 0.42     & 0.07     &  0.81 \\
S55-uM100-IaK  & 3.49             &  -0.009    &0.088           &  0.061     & 0.53     & 0.07     & 0.83   \\
K93-uM40-IaK    & 3.41             &  -0.011   &0.085           &  0.065    & 0.36     & 0.08     & 0.85  \\
K93-uM100-IaK  & 3.41             &  -0.009   &0.084           &  0.066    & 0.38     & 0.08     & 0.86  \\
K01-uM40-IaK    & 3.64             &  -0.009   &0.078           &  0.066      & 0.73     & 0.12     & 0.85 \\
K01-uM100-IaK  & 3.45             &  -0.009    &0.080           &  0.066     & 0.97    & 0.13     & 0.86 \\
S55-uM100-IaK3 & 2.95            &   -0.011   &0.094           & 0.054    & 0.51       & 0.14    & 0.81\\
S55-uM100-IaM  & 3.19             &   -0.008   &0.091           &   0.058    & 0.41     & 0.08  & 0.82\\
S55-uM100-IaGR & 3.58             &   -0.007   &0.087           &   0.062    & 0.49    &  0.08   & 0.85\\
\end{tabular}
\end{center}
\end{table*}

\subsection{Star formation and supernovae rates}

\begin{figure*}
\includegraphics[width=1.0\textwidth]{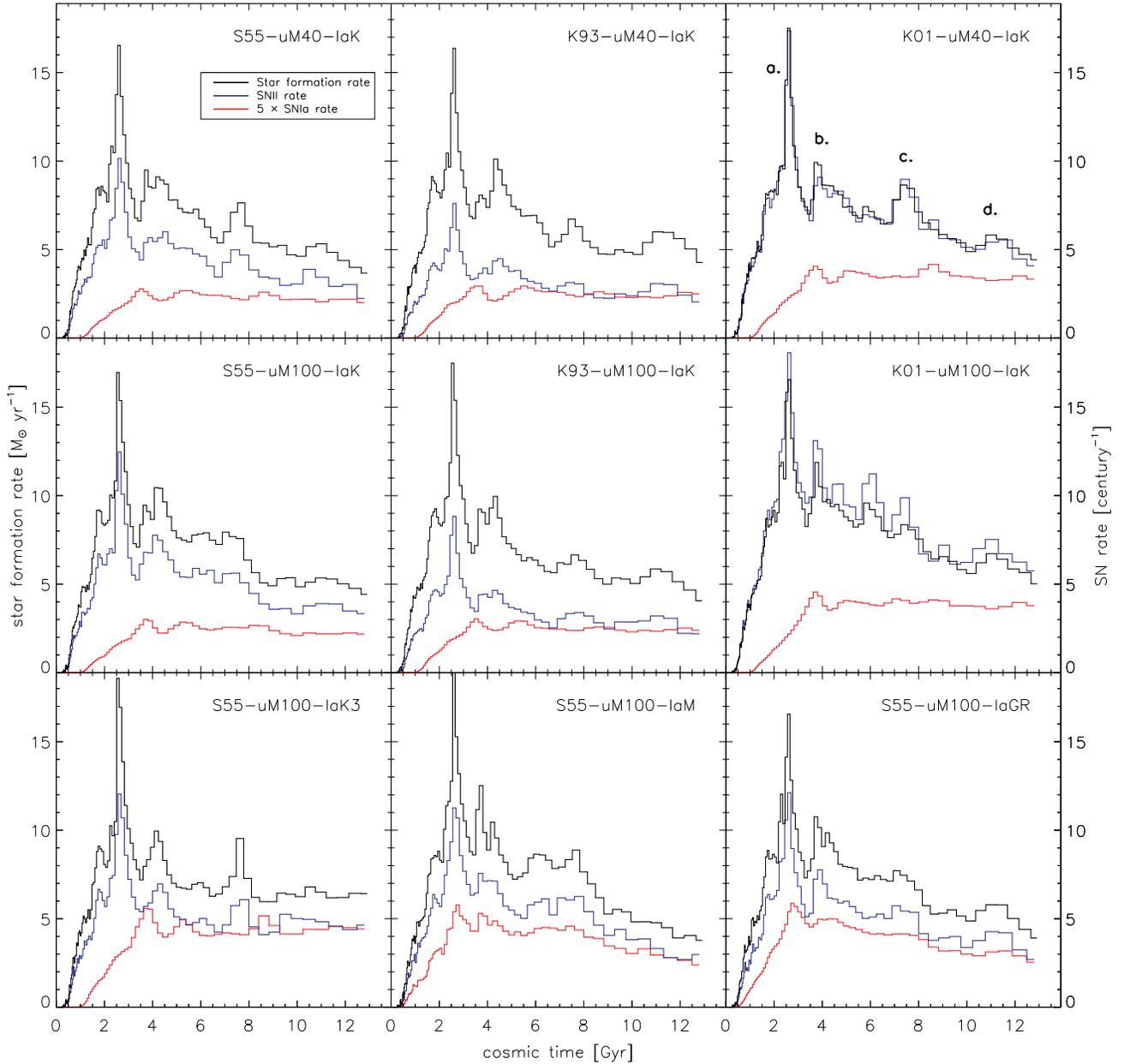}
\caption{Star formation and supernovae rates of each simulation. The black
  line indicates the star formation rate with the scale given on the
  left axes. The SN rates are plotted in blue for SNII and red for
  SNIa. The units are given on the right axes but note that for
  clarity the SNIa rates are scaled up by a factor of 5. The letters in the upper right panel 
  are used to label the four peaks in star formation for reference in \S\ref{streamsec}.}
\label{fig:sncomp}
\end{figure*}

In Fig.~\ref{fig:sncomp} we show the star formation history and SN
rates for each realisation of the galaxy among which is the model published 
previously in \cite{few12}, \emph{K01-uM40-IaK}. A spatial cut has been
employed in all cases to exclude the nearby satellite with the in-plane distance
from the center limited to 15~kpc and distance from the disc plane
limited to 3~kpc. Fig.~\ref{fig:sncomp} demonstrates the expected
behaviour of SNII rates (in blue) as a function of the underlying IMF. Compare
the top two rows of Fig.~\ref{fig:sncomp} which have different IMFs
with the same upper mass limit (40~M$_\odot$ on the top row and
100~M$_\odot$ on the middle row): the SNII rates are greater in
realisations that are more top-heavy (i.e. \emph{K01}). If we now look at the models
with different value of $m_\mathrm{SNII,u}$ by comparing models in the
top row of Fig.~\ref{fig:sncomp} with those immediately below them,
increasing $m_\mathrm{SNII,u}$ increases the SNII rate of the
galaxy. The IMF has a much weaker effect on
the SNIa rate and does not affect the SNIa rate of \emph{IaGR} or
\emph{IaM} models at all, in these cases it is due to the fact that the DTD is 
used independently of the IMF. 

The key difference between realisations using differing SNIa models is also
as expected, the SNIa rates have slightly different magnitudes but more
importantly the onset time of SNIa clearly differs. The appearance of
the first SNIa occurs 0.7~Gyr after the first SNII in realisations
employing \emph{IaK} SNIa schemes while no offset is visible for those
using the \emph{IaGR} and \emph{IaM} schemes. The initial onset of SNIa may
not have a strong impact on any but the most metal-poor stars; however,
the models with earlier SNIa onset times also create global SNIa rates
that have peaks at the same time as the peaks in SNII. This contrasts
with the \emph{IaK} scheme that results in SNIa peaks offset from the
SNII peaks, a point to which we return later. The final distinction to be made is that the single \emph{S55-uM100-IaK3}
realisation gives a much greater SNIa rate and in comparison with
\emph{S55-uM100-IaK} gives an obviously different star formation
history due to the increased thermal feedback. We echo this result by
applying the \emph{IaK3} model to a second IMF (\emph{K01}). Due to
the larger number of massive stars found with a \emph{K01} compared 
with a \emph{S55} IMF the enhancement to the SNIa rate is even higher
in this case. We show the star formation histories, SN rates and
abundance ratio diagrams in Appendix~B.

The reader may notice in Fig.~\ref{fig:sncomp} that the star formation
histories of the realisation look broadly similar as one would expect
for simulations with the same initial conditions, but that the final
peak in star formation (labelled 'd.') is apparently absent from the
panels for \emph{S55-uM100-IaK3} and \emph{S55-uM100-IaM}. These two
runs are characterised by low gas fractions compared with the other
runs. It is probable perturbing satellites
responsible for causing the peaks in star formation are likewise gas
poor. Furthermore, the orbits of the satellites are affected by their
mass and the mass profile of the host galaxy both of which are different for
each realisation and will influence how close the perturbing or
accreted satellite comes to the disc.

We also compare the quantitative SN rates within the virial radius with those observed by 
\cite{mannucci08} for field Sbc/d galaxies:
0.140$^{+0.045}_{-0.035}$~SNuM for SNeIa and
0.652$^{+0.164}_{-0.134}$~SNuM for SNeII.\footnote{SNuM units give the
number of SN per 10$^{10}$~M$_\odot$ stellar mass.} 
We find that most of our SNIa rates are too low, however
  models \emph{K01-uM40-IaK}, \emph{K01-uM100-IaK} and
  \emph{S55-uM100-IaK3} are within the error bars of the
  \cite{mannucci08} observations. SNII rates are consistent with
  observed values for models \emph{S55-uM100-IaK} and
  \emph{K01-M40-IaK} while the majority of other models have SNII
  rates that are too low.  Based on the SN rates alone, the most
  successful model is \emph{K01-uM40-IaK} which is consistent with
  observed values of both SNeII and SNeIa. This underestimation of the
SN rates may at least be partially because they are normalised by
stellar mass; the star formation rate of our models at early times is
very likely too high which will enhance the stellar mass more than the
present day SN rates. Our models also tend to have greater present
day star formation rates than observed (a feature common to many
numerically simulated galaxies with feedback schemes similar to ours)
which would lead to an overestimation of the absolute SNII and SNIa (depending on the
promptness of the DTD) rates. The issue of correctly reproducing the star formation history of the galaxy should be
addressed before drawing firm conclusions about the IMF based on this
kind of analysis.

\subsection{Stellar abundances}

\begin{figure*}
\includegraphics[width=1.0\textwidth]{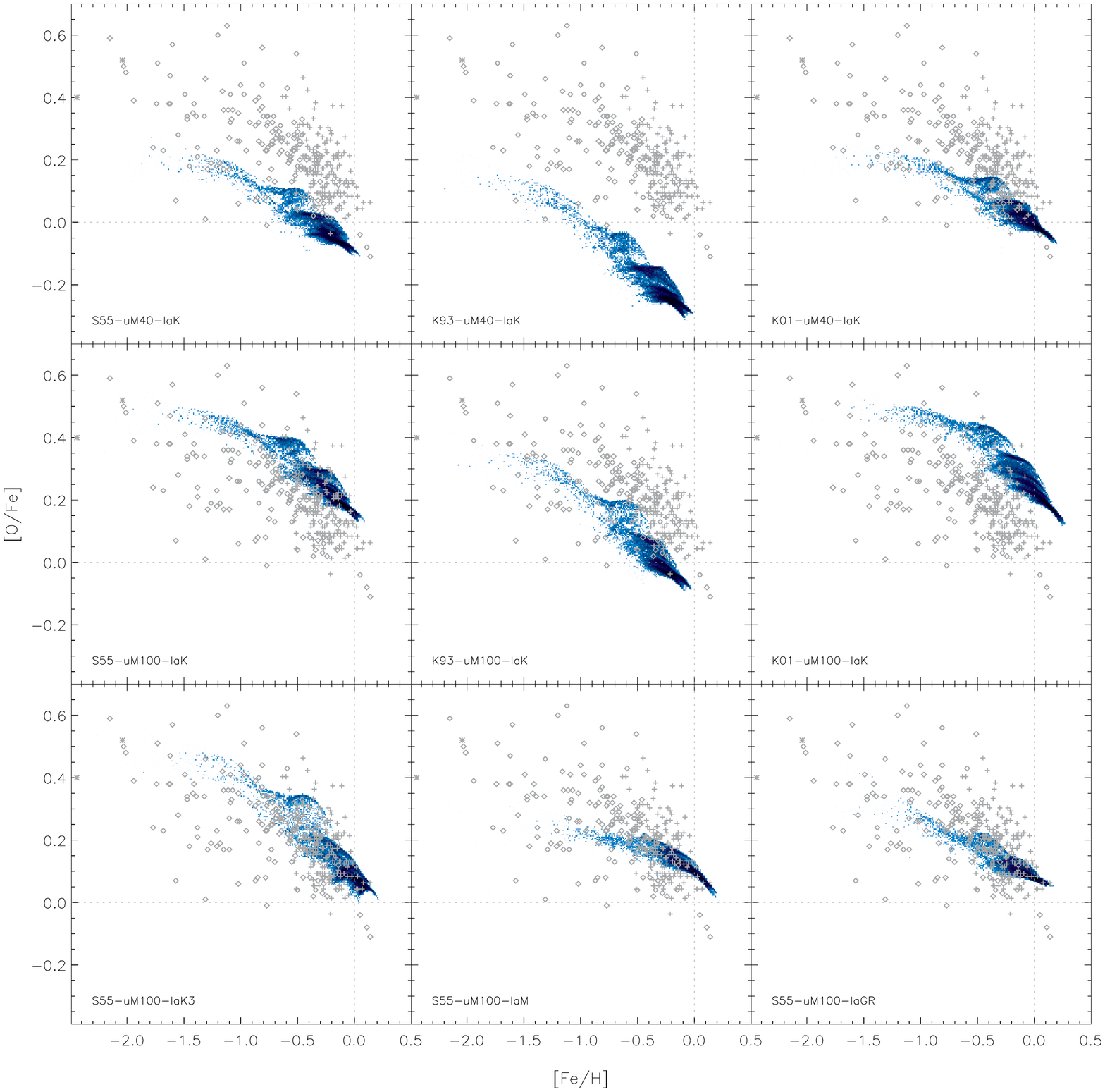}
\caption{Distribution of `solar neighbourhood' stars in the abundance
  plane [O/Fe]-[Fe/H]. Stars are selected from within an annulus of
  galactocentric radius spanning 5--11~kpc and with 
height above (or below) the disc plane of no more than 3~kpc and
placed in abundance ratio bins weighted by mass. Darker colours indicate
a higher relative frequency density. Dotted lines indicate solar abundances.
Observations are plotted in gray, diamonds: thick disc and halo stars \citep{gratton03}, plus
signs: F and G dwarf stars in the thick and thin disc \citep{reddy03},
and asterisks: very metal-poor stars \citep{cayrel04}.}
\label{fig:ofe}
\end{figure*}

Stellar abundance ratios provide a record of the chemical evolution of
the galaxy and are regularly used as a constraint on CEMs; the
ability of our models to recover this distribution is essential. We now
consider the ratios [O/Fe] and [Fe/H] where the square brackets denote
that we are taking the logarithm of the mass ratio of the two elements
normalised to the solar abundance of those elements. We plot the
mass-weighted relative frequency density of stars for these two abundance ratios in
Fig.~\ref{fig:ofe}. This figure tells us how stars are distributed in
the abundance ratio plane, older stars tend to inhabit the low [Fe/H]
part of the figure while younger stars are to be found nearer the
origin (indicated with dotted lines). Also shown in Fig.~\ref{fig:ofe} are the observations of thin disc, thick disc and
halo stars \citep{gratton03, reddy03,cayrel04}. Again we note that \emph{K01-uM40-IaK} is the 
previously published model \citep{few12}. We have re-normalised observational data to the
same zero-point using the solar abundance measurement of
\cite{anders89} to allow a consistent comparison to be made between the simulated 
abundances and the observations that use a variety of solar abundance determinations. 
As these observations are generally limited to
the solar neighbourhood, and in analogy with \cite{calura12}, we have endeavoured to select stars from
the disc at a similar galactocentric radius to the sun, i.e. 5--11~kpc and with 
height above (or below) the disc plane of no more than 3~kpc. It
should be noted that due to the shallow metallicity gradients
of these simulated galaxies the choice of annulus does not affect the
results appreciably.

The majority of the models recover the general qualitative trend with
an initially high [O/Fe] and a `knee' at approximately
[Fe/H]=-1, stars with a great Fe abundance than this have
progressively lower [O/Fe]. It can also be seen that the majority of
stars have these lower [O/Fe] abundances. Two models do not exhibit
this behaviour, \emph{S55-uM100-IaGR} and
\emph{S55-uM100-IaM}, and instead have a far smaller dynamic range in
[O/Fe] while still extending over the same range in [Fe/H], this is explained further in \S\ref{lockstep}.

Fig.~\ref{fig:ofe} also demonstrates the strength of the effect of
varying the IMF for the simulated galaxies. Models with a top-heavy IMF 
have a significantly higher O abundance, e.g. models with \emph{K93} have [O/Fe]
that is offset lower by $\sim$0.2~dex compared with their \emph{K01}
counterparts. Furthermore the mean [O/Fe] is influenced by changes in $m_\mathrm{SNII,u}$ with
[O/Fe] rising by $\sim$0.25~dex when we increase $m_\mathrm{SNII,u}$
from 40~M$_\odot$ to 100~M$_\odot$. Unlike with changes in the IMF slope,
the peak [Fe/H] reached is unaffected due to the very low Fe yield
from stars with masses above 40~M$_\odot$.

We calculate the mean [Fe/H] of the solar neighbourhood analogue for
each of the realisations (for reference, the maximum [Fe/H] visible in Fig.~\ref{fig:ofe} is
a reasonable proxy) and find a similar trend is seen in the mean value of [Fe/H] with the IMF slope as with [O/Fe]; top-heavy 
IMFs give higher mean Fe abundances by around 0.2~dex. Altering $m_{\mathrm{SNII,u}}$ appears 
to have less of an impact, again due to the low Fe content in the ejecta of the most massive stars. 
As expected the enhanced SNIa feedback of \emph{S55-uM100-IaK3} greatly increases 
the quantity of Fe in the galaxy to a similar value as that models
using the top-heavy IMF by \cite{kroupa01} and the 
prompt SNIa models \emph{S55-uM100-IaGR} and \emph{S55-uM100-IaM}. In
Appendix B we show the effect on the \emph{K01} IMF of introducing
the \emph{IaK3} SNIa binary fractions; this leads to a large excess in the Fe content of the galaxy.

It is worth noting features that make this approach different to semi-numerical modelling and one
such feature is seen at [Fe/H]$\approx$-0.5 where [O/Fe] ceases to
decline with increasing [Fe/H] for a short time before resuming its
decline at slightly greater [Fe/H] to its previous trajectory. Several
of these `knees' appear in all the realisations except for the two that have
a low spread in [O/Fe] (\emph{S55-uM100-IaGR} and
\emph{S55-uM100-IaM}). These knees are linked to the episodic 
star formation of the galaxy (which is itself caused by discrete
galaxy mergers) and are described further in the next subsection.

Many of the models do not exhibit a great enough dynamic range in [O/Fe]
to successfully match the [O/Fe] of both Fe-poor and solar Fe stars. Two of
the models \emph{S55-uM100-IaGR} and \emph{S55-uM100-IaM} have already
been noted as failing in this respect but even the \emph{IaK} models
do not have a steep enough slope to successfully match [O/Fe] across
the full range of observations. Only the run with enhanced SNIa
numbers (\emph{S55-uM100-IaK3}) has a steep enough slope to
succesfully match the oxygen abundance across the full range of metallicities.

\subsection{Dissecting abundance space}
\label{streamsec}
\begin{figure}
\includegraphics[width=84mm]{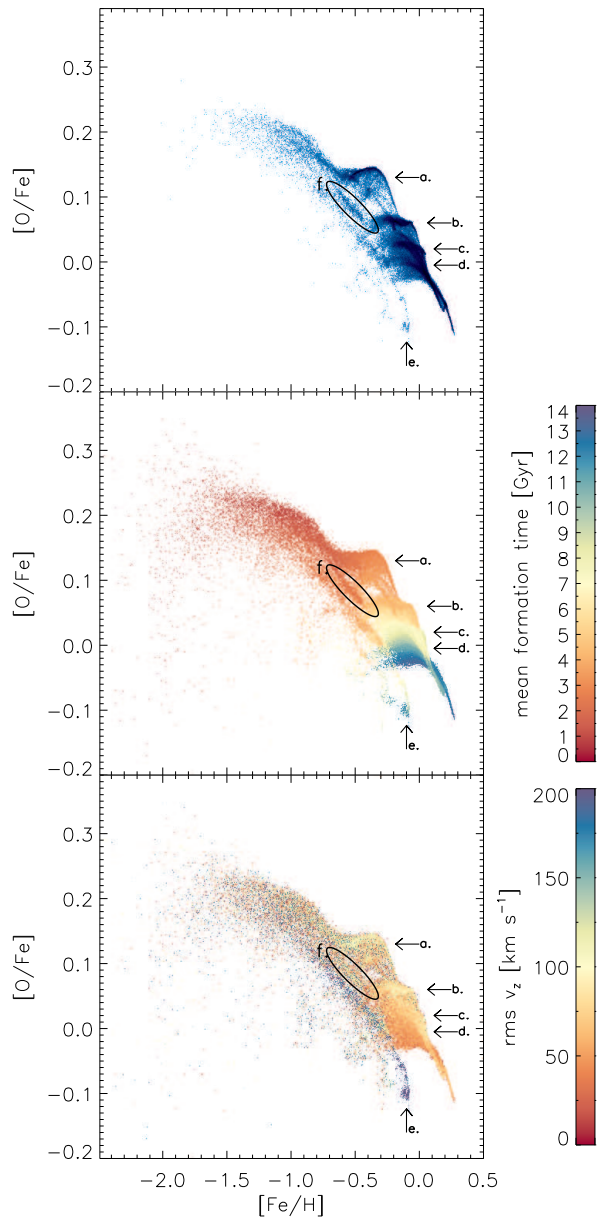}
\caption{Properties of stars in different regions of the [O/Fe]-[Fe/H]
  plane for \emph{K01-uM40-IaK}. We bin the stars according to their 
  abundance ratios [O/Fe] and [Fe/H]. In the top panel we colour bins 
  by the relative mass of stars in each bin (as with Fig.~\ref{fig:ofe}) 
  to highlight four `strata' (labelled a.--d.)
  that are well populated and a fifth feature that has a distinctive
  low [Fe/H] and [O/Fe]. Darker colour denotes a greater mass. The stars encircled and labelled f. do not 
  form in the galaxy as described more fully in the text. 
  In the middle panel we colour bins by the 
  mass-weighted average formation time of star particles that they contain. 
  In the lower panel we colour bins according to the mass-weighted
  root-mean-square velocity of the enclosed star particles.}
\label{fig:streams}
\end{figure}

We now examine the chemical element ratio plot for a single galaxy
to highlight several interesting features of the abundance plane.
In Fig.~\ref{fig:streams}
we bin the stars within the virial radius in one of our realisations (\emph{K01-uM40-IaK}) according to 
their abundance ratios [O/Fe] and [Fe/H]. Briefly we recall that the
schematic behaviour in [O/Fe] vs. [Fe/H] for a galaxy with a monotonic
star formation history is an initially high [O/Fe] plateau extending
from the lowest [Fe/H] up to $\approx$-1 followed by a `knee' at which
point the [O/Fe] declines as [Fe/H] increases until both reach the
solar value. In our galaxies we see multiple knees preceeded by
short-lived plateaus that we will refer to as `strata', these arise
from the bursty star formation history.

In the top panel of Fig.~\ref{fig:streams} the bins
are coloured by the relative mass of stars in each bin (as in
Fig.~\ref{fig:ofe}) to highlight four strata
labelled a.--d. which appear to have a limited evolution in [O/Fe]
even as [Fe/H] spans a range of approximately 0.3~dex. A fifth feature (e.)
has an apparently independent chemical evolution `stream' with
[Fe/H] offset from the majority of other stars by around
0.4~dex (this feature is most clear in the bottom panel of Fig.~\ref{fig:streams}). Henceforth we use the term `stream' to refer to coherent
features in the abundance ratio plane that move diagonally downwards
left-to-right in the same fashion as feature e.
Parallel to e. is a more metal-rich stream labelled f. the origin of 
which we describe later.

In the middle panel of Fig.~\ref{fig:streams} bins are coloured by the
mean formation time of the stars therein, indicating that within each of the strata
a.--d., the stars have a common formation time. We find that these strata
correspond to the four clear peaks in the star formation rate at  2.5~Gyr,
4.0~Gyr, 7.5~Gyr and 11.0~Gyr seen in Fig.~\ref{fig:sncomp} (labelled with the same letters in the upper right panel) during which the SNII rate also
increases to temporarily enhance the O content of the galaxy. These
successive bursts of star formation thus result in distinctive knees
in the top panel of Fig.~\ref{fig:streams} as the mean [Fe/H] of each
strata increases and the mean [O/Fe] decreases since the last episode
of star formation. It is worth noting that from the upper panel of
Fig.~\ref{fig:streams} it is clear that stratum a. exhibits an up-turn
where [O/Fe] actually increases while b. merely has a constant value in
[O/Fe] and c. and d. show increasing ranges of [O/Fe] which is likely
due to the strength of the star formation bursts (and thus the
associated SNII bursts) diminishing with time (see
Fig.~\ref{fig:sncomp}) and the increasing number of SNIa after the
feature with a typical age of 2.5~Gyr. Bear in mind that the rate at
which [O/Fe] declines with [Fe/H] will depend upon the number of SNIa;
sub-grid chemical evolution models with greater SNIa rates will cause
these strata to blur into one another.

\begin{figure}
\includegraphics[width=84mm]{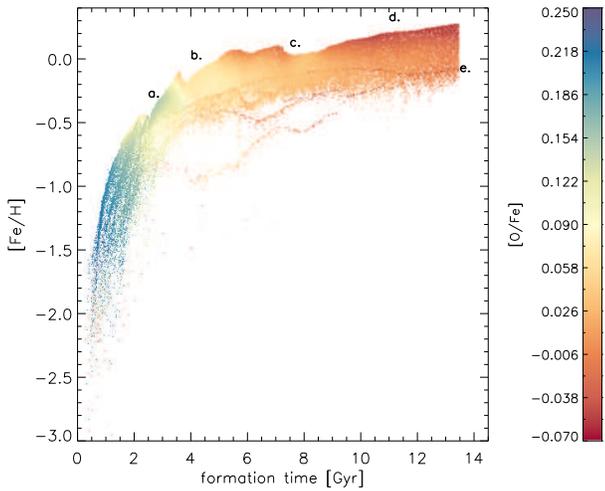}
\caption{The age-metallicity relation of \emph{K01-uM40-IaK}. Each bin in the figure is coloured according 
  to the [O/Fe] of the stars it contains. We use the labels a.--d. again to note the effect of mergers causing a short-term 
  decline in [Fe/H]. The satellite is also visible in
  this figure as a thin line of darker red bins against the background of
  the central galaxy which terminates at label e. The two low-[Fe/H] streams seen in this figure are composed of very few stars and are not attributed to 
  any object in this work.}
\label{fig:agemetstreams}
\end{figure}

In the lower panel of Fig.~\ref{fig:streams} the bins are coloured
according to the root-mean-square vertical velocity (henceforth
\emph{velocity dispersion}) of the stars enclosed. In
\cite{haywood13}, solar neighbourhood stars are separated into thin and thick discs
based on abundance ratios and age, the thick disc is comparable to the
regions labelled as a. and b. here while the thin disc is similar to
c. and d. combined. \cite{haywood13} give the velocity dispersions of
the solar neighbourhood thick and thin disc as 22--50~km~s$^{-1}$ and
9--35~km~s$^{-1}$ respectively. In another study, \cite{adibekyan13}
find mean values of 36~km~s$^{-1}$ and 18~km~s$^{-1}$. These are lower than the typical values of
80--100~km~s$^{-1}$ in regions a./b. and 33-40~km~s$^{-1}$ in regions c./d.
Simulated galaxies are well known as being kinematically hotter due to
resolution effects that heat the particle ensemble \cite[e.g.][]{knebe00,house11}, but we note the
trend is qualitatively similar, i.e. the older and more alpha-rich
disc stars have a higher velocity dispersion. The existence of higher velocity dispersion disc stars that are more
Fe-poor and $\alpha$-rich is consistent with a series of older 
thicker discs, but it must be said that
the vertical density profile appears as a continuous transition to
longer scale heights rather than a distinct set of scale lengths.
The concept of a continuous vertical density distribution that may be interpreted 
as a thin-thick disc division is explored in \cite*{bovy12} though here it must 
be remembered that any simulated thin disc component may be erased due to 
the resolution. The trends of velocity dispersion and age of stars in different regions in 
the abundance plane appear to be qualitatively the same as those found 
by \cite{schoenrich09}.

We find a picture emerging of a series of superposed disc structures
forming with almost discrete ages corresponding to bursts in star
formation. The older the disc is the higher its [O/Fe] and the greater
its velocity dispersion (and consequently the greater the scale height).
This behaviour has been found previously by \cite{bird13} (without the
context of element abundances) where the youngest stars have velocity
dispersions at around 20~km~s$^{-1}$, while older populations have
progressively larger velocity dispersions with the oldest populations having
dispersions ranging from 60--160~km~s$^{-1}$ depending on the
galactocentric radius. The trends of velocity dispersion, age and
abundance ratios are studied in \cite{minchev13}. While the velocity 
dispersion is again also a function of position, stars with
0.14$<$[O/Fe]$<$0.26 have velocity dispersions of 33--51~km~s$^{-1}$ 
while those with -0.1$<$[O/Fe]$<$0.02 are lower at
15--43~km~s$^{-1}$. The same qualitative results are shown by
\cite{stinson13} where the scale height of mono-abundance populations is
considered in place of the vertical velocity dispersion: the
mono-abundance populations with low [O/Fe] have the shortest scale heights.
In each of these works the region of the galaxy considered can have a strong impact on the precise values of velocity
dispersion due to radial and vertical trends, however we see a consistent trend
of older disc populations having greater velocity dispersion
(i.e. physically thicker) that are also $\alpha$-rich and Fe-poor
compared with the youngest population that reside in a thinner, Fe-rich disc.

The feature e. has a large velocity dispersion and highlights the stars
belonging to the nearby satellite at a distance of $\sim$60~kpc that has
a relatively $\alpha$-poor chemical evolution stream independent of
the main galaxy. The satellite may be considered analogous to the
Large Magellanic Cloud whose stars are offset to lower [$\alpha$/Fe]
from Milky Way stars \cite[e.g.][]{johnson06,lapenna12,vdswaelmen13}.

In addition to the appearance of the satellite as feature e. which has
an obviously different position and velocity dispersion there is
another stream indicated by an oval (labelled f.) in the top panel of Fig.~\ref{fig:streams}
which is kinematically indistinct from the rest of the galaxy and yet
follows an abundance ratio stream parallel to the satellite and does
not have the same bumps as the central galaxy. This would suggest that
these stars originate in separate objects that have since merged with
the central galaxy. We tested this hypothesis by extracting the stars
found in this region of the abundance ratio diagram and examining the
location of these stars when they formed. This confirmed that the
stars originate in several separate structures around the same time as
merger activity that resulted in feature a. which later merged with
the main galaxy. The extracted stars are not the only ones that formed
in independent structures, the oldest stars from the structures in question may
also be found mixed with the high-$\alpha$ plateau while the youngest
stars from the structures are mixed amongst those of the high-velocity
dispersion stars belonging to the satellite.

The age-metallicity diagram for \emph{K01-uM40-IaK} is shown in
Fig.~\ref{fig:agemetstreams} with bins that are coloured according to
the [O/Fe] of the stars therein. This figure illustrates clearly the
temporal evolution of the star forming gas phase [Fe/H]. We have once
again marked the merger times/star formation burst times with the
letters a.--d. which shows that the [Fe/H] is briefly diluted before
resuming its Fe enrichment. The [O/Fe] of the stars gradually
decreases from early times to late with only the most Fe-rich stars
extending to sub-solar values.

The colour-coding in Fig.~\ref{fig:agemetstreams} allows one to see a
thin line of relatively [O/Fe] poor stars across the central galaxy
terminating at the point marked `e.' with [Fe/H]=-0.1. This stream
corresponds to the satellite stars marked as e. in
Fig.~\ref{fig:streams}; however, in Fig.~\ref{fig:agemetstreams} the
satellite stream is not discernable without the benefit of
colour-coding to separate it from the central galaxy stars. Two new 
streams are obvious due to their [Fe/H] being much lower than
the central galaxy but are composed of very few stars and we do not
discuss them further. Earlier we discussed the separation of the stars belonging to
feature f. in Fig.~\ref{fig:streams} from those of the central galaxy,
these stars do appear as an independent structure in the
age-metallicity diagram. The stars found in feature f. exist in
Fig.~\ref{fig:agemetstreams} as coeval with the star formation burst
a. at 2.5~Gyr but have similar [Fe/H] and so are not visible as
a distinct structure.

\section{Discussion}
\label{discuss}

\subsection{Oxygen abundance in metal-poor stars}
A key constaint on the models exists in the [O/Fe] of metal-poor stars
as they are born at a time before SNIa or AGB stars have had time to
commence their own enrichment of the ISM, allowing us to examine the
effect of SNII enrichment alone. It is clear in Fig.~\ref{fig:ofe} that only the \emph{uM100} models
(those with $m_\mathrm{SNII,u}$=$100$~M$_\odot$) are successful in
reproducing a high enough value of [O/Fe] for metal-poor stars. 
If the chosen models is correct this indicates 
that there is a need for an initial burst of SNII with masses
up to 100~M$_\odot$ to produce sufficient oxygen in the model galaxy.
This may not however indicate that this upper mass limit is a true representation 
of that found in nature because of the inherent uncertainty in the abundance of 
oxygen and iron in high mass stars. As such this result must be viewed in the context 
of our chosen nucleosynthesis model. 
This uncertainty is extrapolated for models with $m_\mathrm{SNII,u}$$>$40~M$_\odot$ but should have little 
impact on our \emph{uM40} models. The qualitative results are consistent with the analytical results in \cite{gibson98}, i.e. that larger values of
$m_\mathrm{SNII,u}$ generate greater [O/Fe]. \cite{gibson98} does however find
40~M$_\odot$ as the minimum value for $m_\mathrm{SNII,u}$ and notes, as we wish to, that the
[O/Fe] plateau does not place tight constraints on $m_\mathrm{SNII,u}$
due to uncertainties in the yields of massive stars.

Another consideration here is our assumption that all stars
more massive than 8~M$_\odot$ end their lives as SNII; hypernovae, type
Ib SN and type Ic SN may also be important ingredients. \cite{kobayashi11} finds that as many
as 50\% of core collapse supernovae more massive than 20~M$_\odot$ are required to be the more
energetic hypernovae in order to reproduce Zn abundances. This is
something we hope to incorporate in future work, particularly since
the introduction of hypernovae allows the stellar population to
produce stronger feedback energy close to star formation sites with
the effect of reducing stellar mass fractions.

One of our main findings is that the mass range of SNII has a strong
effect on low metallicity [O/Fe] values meaning that the poorly constrained yields of
stars above 40~M$_\odot$ are of particular interest and future
modelling of nucleosynthesis both in massive stars and during the
explosive SN phase are extremely valuable.

\subsection{Lockstep evolution}
\label{lockstep}
We find that the two realisations employing SNIa progenitors models
with more prompt DTDs (\emph{S55-uM100-IaGR} and \emph{S55-uM100-IaM}) 
have a distinctive chemical evolution. SNIa occur much earlier in
these models than in models using the more delayed SNIa model
\emph{IaK}. This can be seen in Fig.~\ref{fig:sncomp} in the
initial onset time of SNIa being earlier in these models which gives
rise to oxygen-poor low metallicity stars in comparison with models
with the equivalent IMF and $m_\mathrm{SNII,u}$. Another aspect of this and one that is
perhaps more important is that throughout
the galaxy's evolution, most of the other models exhibit an offset between
peaks in the SNII rate and in the SNIa rate; this is not so for the
\emph{S55-uM100-IaGR} and \emph{S55-uM100-IaM} simulations where the SNII and SNIa peaks
coincide. This manifests in a `lockstep' evolution of the abundance
ratios shown in Fig.~\ref{fig:ofe} as there is less temporal
separation in the elements produced by SNIa (e.g. Fe) and those
produced by SNII (e.g. O). This is the reason for the narrow
spread and small dynamic range in [O/Fe] in the \emph{IaGR} and
\emph{IaM} simulations. Since the characteristic turnover seen at around [Fe/H]=-1
seen in other models is due to SNIa becoming relevant sources of
enrichment, this feature is not clearly seen in the
\emph{S55-uM100-IaGR} and \emph{S55-uM100-IaM} realisations despite the
overall slope with [Fe/H].

A channel that produces tardy SNIa is described in \cite{yungelson98}
whereby `edge-lit detonation' occurs in the accreted helium of a C/O
WD at sub-Chadrasekhar masses producing a DTD that is similar to that
of \cite{kobayashi98} but without a gap.
The requirement for tardy SNIa to reproduce the knee in the abundance
plane has been demonstrated by \cite{kobayashi98, kobayashi00} where a
double-degenerate SNIa scenario \citep{tutukov94} is compared with the single degenerate
model we use in this work (\emph{IaK}) finding that the
double-denerate scenario produced a weak knee, with the single
degenerate doing little better. The best model employed the
single degenerate DTD with an addition metallicity limit that
prevented systems with [Fe/H]$<$-1.1 from producing SNIa. The
metallicity limit is required to represent the effect of
the metal content in optically thick winds that transfer mass from the companion
to the WD \citep{hachisu99}. We find that the existance of a metallicity limit has
little effect on the distribution of stars in the abundance plane, but
this does not give evidence for or against a metallicity bias in SNIa progenitors.
Evidence counter to the concept of metallicity limited SNIa may be
found in observed SNIa in low metallicity environments
\citep{strolger02,quimby06,prieto08} but the existence of multiple SNIa
channels may provide an explanation for this inconsistency. 

Semi-numerical models are used to compare different DTDs in
\cite{matteucci09}, among them that of \cite{matteucci01} which is
very similar to the \emph{IaGR} used in this work, \emph{IaM} and a
more tardy model with no prompt component (the first systems explode
after 2.5$\times$10$^8$~yr, with the peak at 3~Gyr) proposed by
\cite{strolger04}.\footnote{Note the errata to \cite{strolger04}.} In terms of fitting to observed abundance ratios
the two DTDs with prompt components produce abundance ratios falling
within observed limits, whereas the tardy
\cite{strolger04} DTD produces a pronounced knee (perhaps too pronounced)
with a plateau at [O/Fe] values above those observed at low
metallicity, in each case the models are normalised to solar
abundances. In \cite{matteucci01} there is a clear failure of their \emph{IaGR}-like DTD to 
reproduce the high-[O/Fe] plateau up to [Fe/H]=-1 exhibiting as it
does the same gradual decline that we find.

\subsection{Type Ia supernova explosion efficiency}

It is well known from analytical formulations of chemical evolution
that the SNII/SNIa ratio is influenced by the typical time-scale of
star formation \cite[e.g.][]{matteucci86, tsujimoto95}. We note that, as with many simulated galaxies, the
star formation rate of the galaxy presented in this work is too high
at low redshift. Excessive star formation at late times gives rise to
excessive SNII rates and makes the slope of abundance ratios too
shallow above the [Fe/H]$\approx$-1 knee and the additional production of
alpha-elements from merger-induced star formation bursts likely
worsens the problem. It is clear that the decline in [O/Fe] in the high-metallicity 
regime is not great enough in the \emph{IaK} models, perhaps due to the poorly reproduced star formation
history. In spite of this we wish to demonstrate how the assumed
number of SNIa per unit mass of stars formed influences chemical evolution in a numerical context
and to do so we increase the fraction of systems giving rise to SNIa in Eq.~\ref{eq:IaK}, 
factors $b_\mathrm{RG}$ and $b_\mathrm{MS}$, by a factor of 1.8 (these models are labelled with \emph{IaK3}).

The enhancement to the SNIa progenitor fraction in the \emph{IaK3}
realisations is adequate to steepen the decline in [O/Fe] with
increasing [Fe/H] and match the observed slope. We would like to 
note that enhancements in $\alpha$-element production may come from the 
late excess in star formation and also from the exaggerated effect of gas-rich 
galaxy mergers inducing large bursts in star formation which will result in a slower 
decline (or even an increase) in [O/Fe] towards higher values of [Fe/H].
This means that in the absence of these systematic effects the models that reproduce the
observed distribution in abundance space will be too steep and as such
these models should be viewed as representing upper limits to the SNIa
progenitor fraction. To simplify the following discussion we borrow
the concept of `explosion efficiency' from \cite{madau98}, that is the
fraction of stars with masses between 3-8~M$_\odot$ that give rise to
SNIa. \cite{maoz08} compiles
numerous estimates of this fraction that range between 2-40\%. 
Estimates calculated from abundance ratios tend to be larger at 11-40\%
\citep{dePlaa07, maoz08} as does the value given by
\cite{mannucci05} from examination of the number of SNIa
relative to core-collapse supernovae in star forming galaxies
(8-15\%). In contrast more direct measurements of the SNIa rate per
unit star formation rate \citep[e.g.][]{dahlen04,scannapieco05b,sullivan06} 
and per unit mass \citep[e.g.][]{mannucci05,sharon07,mannucci08} 
are lower at 0.8-18\%. In our \emph{IaK} realisations the fraction is 4.7\% while the \emph{IaK3}
have 8.46\% which is closer to the 7\% used for the simulated galaxies
in \cite{kobayashi11} where a reasonable agreement for many abundance
ratios was found with a similar implementation to our SNIa
feedback. 

While it is beyond the scope of this work, a time varying IMF that
steepens with metallicity \citep[e.g.][]{suda13} also has the potential to enhance Fe and
reduce O production.  Given the controversial nature of
variable IMFs we leave this for a more detailed study.

\subsection{Abundance strata}

We have examined the properties of stars in different regions of the
[O/Fe]-[Fe/H] plane to determine how chemical properties may be used
to disentangle the origin of those stars. We find that stars may be
traced to their formation within independent (prior to accretion by the
main galaxy) dark haloes or to individual bursts of star formation
induced by mergers.

The decomposition highlights the importance of bursty star formation
histories by showing that there are four bands with a narrow spread in
[O/Fe] but with a range of [Fe/H]. The ages of stars in these bands
are well matched to the time at which the star formation rate
increases temporarily during mergers. The rise in star formation
creates an enhancement in the number of stars with the particular
elemental abundances of the gas at approximately that time. We also
see a temporary enhancement in the SNII rate since it closely follows the SFR. The O produced by the SNII during the star
formation burst shifts the stars to higher [O/Fe] values, effectively
arresting the general trend of declining [O/Fe] with time. We note
that similar behaviour will be observed regardless of the cause of the
increased star formation rate as in \cite{brook12} where only two
abundance strata are seen and are described as an $\alpha$-rich thick
disc and a thin disc population. In this scenario the two strata are delineated by a hiatus in
star formation while the [Fe/H] naturally declines; the authors do not attribute the 
hiatus to any particular event.

We have also examined the vertical velocity dispersion of stars in
each region of the abundance plane. We find that the abundance strata
at lower [O/Fe] (which are composed of younger stars) have lower
velocity dispersions. Since the abundance strata are distinct one
might view these as a series of nested disc populations where the
[Fe/H] increases, the [O/Fe] decreases and the velocity dispersion
(and thus characteristic scale height) increases when comparing older with younger populations.
The distinctiveness of the strata depends upon the strength and
duration of the relevant star formation bursts, if our simulated star formation histories 
are more exaggerated than in reality then our strata will be likewise
exaggerated. Measuring the variations of star formation rate in real
galaxies is difficult but Weisz et al. (2012) find variations that are
less than a factor of ten for galaxies with a comparable mass to our
simulated ones. In that context the ratio of the maximum to the accompanying minimums in star formation rate
for the galaxies in this work are rather low at $\sim$2.

Within the Milky Way a separation of around 0.2~dex in [O/Fe] of two 
stellar populations is seen (Bensby et al. 2005; Reddy et al. 2006; Fuhrmann 2008),
consistent with our findings for \emph{K01-uM40-IaK} (see Fig.~\ref{fig:streams}). This separation may be a consequence of a
hiatus in star formation around 6~Gyr ago that separates the thick
disc formation from that of the thin disc.
Conversely, \cite{schoenrich09} find that even a monotonically
decreasing star formation history can produce this kind of bimodality
in [O/Fe] through the simple fact that after initially forming an
$\alpha$-rich thick disc, the initial onset of SNIa enrichment forces
the [O/Fe] to evolve downward rapidly before settling into a thin disc
steady state. A chemically distinct thick disc is also found in simulations 
by \cite{brook12b} and \cite{gibson13} where the separation of thick and thin disc 
in [O/Fe] is 0.1~dex. This is superficially similar to the results presented here except that 
both of those works find only two offset chemical evolution tracks compared with the three or 
four seen in our galaxy, this is likely due to the more discrete assembly history of our galaxy.

\subsection{Abundance streams}

In addition to the abundance strata, we see another feature in the
abundance plane that we refer to as `streams'. The main galaxy
stellar abundance simply follows the classical form of a high [O/Fe] 
plateau followed by a downturn towards higher metallicities. There is
also an obviously independent stream offset to lower metallicity (labelled e. in 
Fig.~\ref{fig:streams}). Only
the most metal-rich stars have clear and distinct separation from the
main trend but they are more easily separable by age since
stars in the stream are older for a given [O/Fe] than those in the
central galaxy. This particular stream also has a much higher
velocity dispersion than the other galaxy stars. While this stems 
from the fact that these stars belong to a geographically separate 
satellite object, we note that the stream would remain recognisable
even if this satellite had merged at some time in the past. 
This satellite might be considered analogous to the Large
Magellanic Cloud which follows a chemical evolution that is parallel
but offset from that of the Milky Way (Johnson et al. 2006; Lapenna et al. 2012; Van der Swaelmen et al. 2013).

To strengthen our claim that an accreted site of star formation may be 
distinguished from the main galaxy using abundance properties we selected stars from 
another region that remains distinct in the abundance plane (feature f. in Fig.~\ref{fig:streams})
and traced the position of their birth relative to the central
galaxy. While there is contamination of this region by stars that
formed within the central galaxy, the majority are accreted stars born
in several low mass satellites. The stars are now kinematically and spatially
indistinguishable from the galaxy proper but of course retain a distinct 
chemical evolution.

\subsection{Abundance space substructure of other realisations}

We have presented an analysis of the substructure of the abundance plane 
for one of our realisations, \emph{K01-uM40-IaK}. The substructures we have 
noted also exist in the other simulations presented here. This is clearly seen 
in Fig.~\ref{fig:ofe}, but the substructures are slightly less
obvious because this figure contains only stars in the solar
neighbourhood whilst Fig.~\ref{fig:streams} plots all stars in the
galaxy. Comparing the panels in Fig.~\ref{fig:ofe}, the top and middle rows in particular have qualitatively similar 
appearance despite variations in the abundance ratios. \emph{S55-uM100-IaK3} also has an 
obvious second knee, although substructure at lower [O/Fe] is less clear, this is because
the CEM gives rise to only three star formation bursts that are more short-lived than in 
other realisations. As previously discussed, the final two simulations of this galaxy, 
\emph{S55-uM100-IaM} and \emph{S55-uM100-IaGR} have a less clear knee due to the prompt 
onset of SNIa feedback, the slightly less prompt SNIa in \emph{S55-uM100-IaGR} do allow a 
small second knee to form (the bottom right panel of Fig.~\ref{fig:ofe}). This means that 
knees features are less obvious in chemical space which makes the decomposition described 
in \S\ref{streamsec} more difficult. Decomposition of the chemical properties is not helped 
by the smaller dynamic range of these realisations as it means the strata overlay one another.

\section{Conclusions}
\label{conc}

In this work we have implemented additional sources of energetic and
nucleosynthetic feedback within the AMR framework of the \textsc{ramses}
cosmological simulation code. We use the flexibility of the new code
to make changes to the sub-grid SSP model for resimulations of an L$\star$ disc galaxy.
We make changes to the IMF, to the upper mass limit of SNII
progenitors and to the DTD of SNIa. 

The average scale lengths of these galaxies is 3.39~kpc and metallicity gradients that, while shallow, are
within observed limits. While the range of metallicity gradient
reproduced is also narrow compared to observed disc galaxies we do see
hints that runs employing SSPs producing a greater number of SNII
have flatter gradients. It should be made plain, however, that the
flatness of the gradients has much to do with the size of the region
over which feedback is distributed (i.e. it may be a resolution issue)
as well as the star formation distribution and
we are not suggesting that models should make use of less SNII
feedback to reproduce steeper metallicity gradients. We also find that
the metallicity distribution of the galaxy is narrow compared to
observations but with a larger number of stars in the low metallicity
tail as found by \cite{calura12} for other simulated galaxies.

All of the runs have quite similar baryon fractions meaning that
changes in the feedback strength and metallicity of gas in the halo (of the
magnitude produced by our assumptions) do not seem to result in changes 
to the halo baryon fraction unless these effects contrive to cancel one another.
Our variations in the sub-grid physics do seem to affect the
galaxy formation efficiency with changes in the stellar mass
fraction of up to 10\% pointing to changes in the way baryons are redistributed by 
feedback. Since the mass returned to
the ISM by the stellar particles depends upon the IMF, and we wish to
gauge the efficiency with which a realisation forms stars rather than
the efficiency with which stellar mass is retained we calculate the
`mass-formed fraction' which is essentially a normalised integration
of the star formation history. In this case we note that the
realisations with more top-heavy IMFs (those that return more gas to the
ISM on shorter time-scales) are more efficient at forming stars even if
the stellar mass fraction is not the greatest. We attribute this to the 
efficient replenishment of the gas reservoir by top-heavy IMFs.

The star formation rate of our simulated galaxies peaks in the first 3~Gyr and subsequent 
bursts of star formation correspond to mergers. The SNII rate naturally traces 
the star formation history and increasing the upper mass limit of SNII progenitors 
naturally leads to increased SNII rates, as does the use of a more top-heavy IMF.
The SNIa rate is not strongly dependent on the IMF, but the choice of
DTD does make an impact. The only model that satisfactorily reproduces
observed SN rates is \emph{K01-uM40-IaK} while all models but one (\emph{S55-uM100-IaK}) are within
error bars for the ratio of  SNII/SNIa rates. 
DTDs with a more prompt onset result in SNIa rates that peak close to peaks in the star formation 
and SNII rates. The prompt onset SNIa DTDs allow a greater amount of Fe-enrichment 
in the first Gyr of evolution. Increasing the number of SNIa in the \emph{IaK3} (compared with the 
\emph{IaK}) model enhances the amount of thermal feedback and this seems to have a strong impact on the concentration 
of the galaxy. The additional thermal feedback seems to heat the gas around the disc to a temperature 
at which it cools efficiently leading to a concentrated disc with a high star formation rate. 
This is more a function of the thermal mode of feedback employed for SNIa than the efficiency 
of SNIa themselves and in future studies we will explore a kinetic feedback mode for all channels.
The concentration of this galaxy disc has an impact on the galaxy properties, 
i.e. that it has a shorter scale length and a steeper metallicity
gradient.  The strongly dissipative mode of formation seen in this
realisation may be ameliorated with the use of a 
`delayed cooling' method as found in \cite{agertz13} and implemented in \textsc{ramses} by \cite{teyssier13}.

We see trends in the abundance ratio plane of all galaxies that qualitatively match
observed behaviours. Comparing the realisations with one another we
see changes in the oxygen and iron abundances that are also as
expected: runs with more top-heavy IMFs have greater [O/Fe] values
whether caused by the high mass slope or by the SNII upper mass limit.
The reader should note that the nucleosynthesis models we use have a
strong impact on our results. The [O/Fe] of low metallicity stars in
particular is defined by the assumed nucleosynthetic processes in
massive stars, a regime in which we have to extrapolate the results of
\cite{ww95} to masses greater than 40~M$_\odot$ after halving the iron yields.

Our most tardy SNIa model (\emph{IaK}) is successful at reproducing
the turnover at [Fe/H]$\approx$-1 but we find that the more prompt
SNIa DTDs lead to SNIa rates that peak too soon after star formation
and do not allow a relatively SNIa-free phase of enrichment in which the low
metallicity plateau can be established. 

The slope of the [O/Fe]-[Fe/H] relation for stars with [Fe/H]$>\approx$-1
is set partly by the star formation history but we find it can be
controlled through changes in the binary fraction. The
\emph{IaK} models have binary fractions that are too low to match
the slope and so we also present a run with a greater number of SNIa
per unit stellar mass (\emph{S55-uM100-IaK3}) that succesfully reproduces the
slope at high metallicities. The model presented here that uses this
enhanced SNIa feedback is coupled with a Salpeter IMF and a SNII upper
mass limit of 100~M$_\odot$ that reproduces the low metallicity [O/Fe] plateau: this model is
thus our most successful when judged by the abundance ratios alone. 

A final note on the constraints placed on the chemical evolution is that
our conclusions are based on the star formation history of the galaxy
within a simulation. We ask the reader to bear in mind that since our
star formation history is excessive at late times compared to the
Milky Way, the $\alpha$-element production is likewise too high meaning that we may
not need to invoke such high SNIa binary fractions to match the
observed slope in abundance space. Thus we are brought once again to
the perennial issue of correctly regulating star formation.

In this work we have used one of our realisations to consider the
physical properties of stars in different regions of the abundance
ratio plane. We note several features of interest; a succession of
strata with lessening [O/Fe] that are produced by consecutive star
formation bursts, stars belonging to a satellite that traces a
parallel chemical evolution stream offset to lower metallicities and
a second parallel stream with a smaller offset but a far smaller range
in metallicity. We demonstrate that the strata arise during bursts of
star formation and receive enhancements to their [O/Fe] from SNII in
stellar particles formed during those bursts. In the final timestep of
the simulation we find that the younger the abundance strata, the less
kinematically hot it is. This leads us to view the strata as
chemically identified thick discs, or rather a nested series of ageing
and thickening discs.

The satellite retains its individual identity as of the end of the
simulation and is more metal poor than the central galaxy. It also
does not experience any of the upturns in [O/Fe] as the central galaxy
does during mergers. The second stream (that with a smaller offset
from the galaxy proper) cannot be easily distinguished geographically or
kinematically but when stars in the stream are selected they can be
traced back to their birth locations which are in several low mass
accreted bodies.

It is clear that the level of substructure available in abundance
space is apt for decomposing the galaxy into the components from which
it was assembled. If these structures are indeed realistic and not exaggerated 
by numerical effects then they are something which is of particular interest in the era
of large spectroscopic and kinematic surveys. In future versions of our code we 
plan to introduce newer nucleosynthesis models that have become 
available \citep[e.g.][]{doherty10,pignatari13} which cover a greater
range of progenitor masses and are more self-consistent with one
another. We also hope to include type Ib and type Ic SN and hypernovae
into our chemical evolution models as well as improving on the
energetic aspect of stellar feedback.

\section{Acknowledgments}

CGF thanks Clare Dobbs for suggestions that improved the manuscript and 
acknowledges the support of the Science \& Technology Facilities
Council (ST/F007701/1) and funding from the European Research Council
for the FP7 ERC starting grant project LOCALSTAR. 
SC acknowledges support from the BINGO Project (ANR-08-BLAN-0316-01) and the CC-IN2P3 
Computing Center (Lyon/Villeurbanne, France), a partnership between CNRS/IN2P3 and CEA/DSM/Irfu.
Computing resources were provided by the UK National Cosmology Supercomputer (COSMOS), the University of 
Central Lancashire's HPC facility and the HPC resources of CINES under the allocation 2012-c2012046642 and 2013-x2013046642 made by GENCI. 
LMD acknowledges support from the Lyon Institute of Origins under
grant ANR-10-LABX-66. We also thank an anonymous referee for comments
which improved the paper.

\bibliographystyle{mn2e}
\bibliography{ramsesch}

\begin{thebibliography}{135}
\expandafter\ifx\csname natexlab\endcsname\relax\def\natexlab#1{#1}\fi

\bibitem[{{Abadi} {et~al}\mbox{.}(2003){Abadi}, {Navarro}, {Steinmetz}, \&
  {Eke}}]{abadi03}
{Abadi} M.~G., {Navarro} J.~F., {Steinmetz} M., {Eke} V.~R., 2003, ApJ, 597, 21

\bibitem[{{Adibekyan} {et~al}\mbox{.}(2013){Adibekyan}, {Figueira}, {Santos},
  {Hakobyan}, {Sousa}, {Pace}, {Delgado Mena}, {Robin}, {Israelian}, \&
  {Gonz{\'a}lez Hern{\'a}ndez}}]{adibekyan13}
{Adibekyan} V.~Z. {et~al.}, 2013, A\&A, 554, A44

\bibitem[{{Agertz} {et~al}\mbox{.}(2013){Agertz}, {Kravtsov}, {Leitner}, \&
  {Gnedin}}]{agertz13}
{Agertz} O., {Kravtsov} A.~V., {Leitner} S.~N., {Gnedin} N.~Y., 2013, ApJ, 770,
  25

\bibitem[{{Agertz} {et~al}\mbox{.}(2007){Agertz}, {Moore}, {Stadel}, {Potter},
  {Miniati}, {Read}, {Mayer}, {Gawryszczak}, {Kravtsov}, {Nordlund}, {Pearce},
  {Quilis}, {Rudd}, {Springel}, {Stone}, {Tasker}, {Teyssier}, {Wadsley}, \&
  {Walder}}]{agertz07}
{Agertz} O. {et~al.}, 2007, MNRAS, 380, 963

\bibitem[{{Anders} \& {Grevesse}(1989)}]{anders89}
{Anders} E., {Grevesse} N., 1989, Geochimica et Cosmochimica Acta, 53, 197

\bibitem[{{Arnett}(1978)}]{arnett78}
{Arnett} W.~D., 1978, ApJ, 219, 1008

\bibitem[{{Bekki} \& {Meurer}(2013)}]{bekki13}
{Bekki} K., {Meurer} G.~R., 2013, ApJl, 765, L22

\bibitem[{{Bensby} {et~al}\mbox{.}(2005){Bensby}, {Feltzing}, {Lundstr{\"o}m},
  \& {Ilyin}}]{bensby05}
{Bensby} T., {Feltzing} S., {Lundstr{\"o}m} I., {Ilyin} I., 2005, A\&A, 433,
  185

\bibitem[{{Binney}, {Gerhard} \& {Silk}(2001){Binney}, {Gerhard}, \&
  {Silk}}]{binney01}
{Binney} J., {Gerhard} O., {Silk} J., 2001, MNRAS, 321, 471

\bibitem[{{Bird} {et~al}\mbox{.}(2013){Bird}, {Kazantzidis}, {Weinberg},
  {Guedes}, {Callegari}, {Mayer}, \& {Madau}}]{bird13}
{Bird} J.~C., {Kazantzidis} S., {Weinberg} D.~H., {Guedes} J., {Callegari} S.,
  {Mayer} L., {Madau} P., 2013, ApJ, 773, 43

\bibitem[{{Bovy}, {Rix} \& {Hogg}(2012){Bovy}, {Rix}, \& {Hogg}}]{bovy12}
{Bovy} J., {Rix} H.-W., {Hogg} D.~W., 2012, ApJ, 751, 131

\bibitem[{{Brook} {et~al}\mbox{.}(2011){Brook}, {Governato}, {Ro{\v s}kar},
  {Stinson}, {Brooks}, {Wadsley}, {Quinn}, {Gibson}, {Snaith}, {Pilkington},
  {House}, \& {Pontzen}}]{brook11}
{Brook} C.~B. {et~al.}, 2011, MNRAS, 415, 1051

\bibitem[{{Brook} {et~al}\mbox{.}(2012{\natexlab{a}}){Brook}, {Stinson},
  {Gibson}, {Ro{\v s}kar}, {Wadsley}, \& {Quinn}}]{brook12c}
{Brook} C.~B., {Stinson} G., {Gibson} B.~K., {Ro{\v s}kar} R., {Wadsley} J.,
  {Quinn} T., 2012{\natexlab{a}}, MNRAS, 419, 771

\bibitem[{{Brook} {et~al}\mbox{.}(2012{\natexlab{b}}){Brook}, {Stinson},
  {Gibson}, {Wadsley}, \& {Quinn}}]{brook12}
{Brook} C.~B., {Stinson} G., {Gibson} B.~K., {Wadsley} J., {Quinn} T.,
  2012{\natexlab{b}}, MNRAS, 424, 1275

\bibitem[{{Brook} {et~al}\mbox{.}(2012{\natexlab{c}}){Brook}, {Stinson},
  {Gibson}, {Kawata}, {House}, {Miranda}, {Macci{\`o}}, {Pilkington}, {Ro{\v
  s}kar}, {Wadsley}, \& {Quinn}}]{brook12b}
{Brook} C.~B. {et~al.}, 2012{\natexlab{c}}, MNRAS, 426, 690

\bibitem[{{Calura} {et~al}\mbox{.}(2012){Calura}, {Gibson}, {Michel-Dansac},
  {Stinson}, {Cignoni}, {Dotter}, {Pilkington}, {House}, {Brook}, {Few},
  {Bailin}, {Couchman}, \& {Wadsley}}]{calura12}
{Calura} F. {et~al.}, 2012, MNRAS, 427, 1401

\bibitem[{{Calura} \& {Menci}(2009)}]{calura09}
{Calura} F., {Menci} N., 2009, MNRAS, 400, 1347

\bibitem[{{Calura} {et~al}\mbox{.}(2010){Calura}, {Recchi}, {Matteucci}, \&
  {Kroupa}}]{calura10}
{Calura} F., {Recchi} S., {Matteucci} F., {Kroupa} P., 2010, MNRAS, 406, 1985

\bibitem[{{Carbon} {et~al}\mbox{.}(1987){Carbon}, {Barbuy}, {Kraft}, {Friel},
  \& {Suntzeff}}]{carbon87}
{Carbon} D.~F., {Barbuy} B., {Kraft} R.~P., {Friel} E.~D., {Suntzeff} N.~B.,
  1987, PASP, 99, 335

\bibitem[{{Carigi}(1994)}]{carigi94}
{Carigi} L., 1994, ApJ, 424, 181

\bibitem[{{Cayrel} {et~al}\mbox{.}(2004){Cayrel}, {Depagne}, {Spite}, {Hill},
  {Spite}, {Fran{\c c}ois}, {Plez}, {Beers}, {Primas}, {Andersen}, {Barbuy},
  {Bonifacio}, {Molaro}, \& {Nordstr{\"o}m}}]{cayrel04}
{Cayrel} R. {et~al.}, 2004, A\&A, 416, 1117

\bibitem[{{Chabrier}(2003)}]{chabrier03}
{Chabrier} G., 2003, ApJl, 586, L133

\bibitem[{{Chiappini}, {Matteucci} \& {Gratton}(1997){Chiappini}, {Matteucci},
  \& {Gratton}}]{chiappini97}
{Chiappini} C., {Matteucci} F., {Gratton} R., 1997, ApJ, 477, 765

\bibitem[{{Chieffi} \& {Limongi}(2004)}]{chieffi04}
{Chieffi} A., {Limongi} M., 2004, ApJ, 608, 405

\bibitem[{{Chiosi} \& {Caimmi}(1979)}]{chiosi79}
{Chiosi} C., {Caimmi} R., 1979, A\&A, 80, 234

\bibitem[{{Dahlen} {et~al}\mbox{.}(2004){Dahlen}, {Strolger}, {Riess},
  {Mobasher}, {Chary}, {Conselice}, {Ferguson}, {Fruchter}, {Giavalisco},
  {Livio}, {Madau}, {Panagia}, \& {Tonry}}]{dahlen04}
{Dahlen} T. {et~al.}, 2004, ApJ, 613, 189

\bibitem[{{de Plaa} {et~al}\mbox{.}(2007){de Plaa}, {Werner}, {Bleeker},
  {Vink}, {Kaastra}, \& {M{\'e}ndez}}]{dePlaa07}
{de Plaa} J., {Werner} N., {Bleeker} J.~A.~M., {Vink} J., {Kaastra} J.~S.,
  {M{\'e}ndez} M., 2007, A\&A, 465, 345

\bibitem[{{Doherty} {et~al}\mbox{.}(2010){Doherty}, {Siess}, {Lattanzio}, \&
  {Gil-Pons}}]{doherty10}
{Doherty} C.~L., {Siess} L., {Lattanzio} J.~C., {Gil-Pons} P., 2010, MNRAS,
  401, 1453

\bibitem[{{Dubois} \& {Teyssier}(2008)}]{dubois08}
{Dubois} Y., {Teyssier} R., 2008, A\&A, 477, 79

\bibitem[{{Edvardsson} {et~al}\mbox{.}(1993){Edvardsson}, {Andersen},
  {Gustafsson}, {Lambert}, {Nissen}, \& {Tomkin}}]{edvardsson93}
{Edvardsson} B., {Andersen} J., {Gustafsson} B., {Lambert} D.~L., {Nissen}
  P.~E., {Tomkin} J., 1993, A\&As, 102, 603

\bibitem[{{Ferland} {et~al}\mbox{.}(1998){Ferland}, {Korista}, {Verner},
  {Ferguson}, {Kingdon}, \& {Verner}}]{ferland98}
{Ferland} G.~J., {Korista} K.~T., {Verner} D.~A., {Ferguson} J.~W., {Kingdon}
  J.~B., {Verner} E.~M., 1998, PASP, 110, 761

\bibitem[{{Few} {et~al}\mbox{.}(2012{\natexlab{a}}){Few}, {Courty}, {Gibson},
  {Kawata}, {Calura}, \& {Teyssier}}]{few12}
{Few} C.~G., {Courty} S., {Gibson} B.~K., {Kawata} D., {Calura} F., {Teyssier}
  R., 2012{\natexlab{a}}, MNRAS, L463

\bibitem[{{Few} {et~al}\mbox{.}(2012{\natexlab{b}}){Few}, {Gibson}, {Courty},
  {Michel-Dansac}, {Brook}, \& {Stinson}}]{few12b}
{Few} C.~G., {Gibson} B.~K., {Courty} S., {Michel-Dansac} L., {Brook} C.~B.,
  {Stinson} G.~S., 2012{\natexlab{b}}, A\&A, 547, A63

\bibitem[{{Fran{\c c}ois} {et~al}\mbox{.}(2004){Fran{\c c}ois}, {Matteucci},
  {Cayrel}, {Spite}, {Spite}, \& {Chiappini}}]{francois04}
{Fran{\c c}ois} P., {Matteucci} F., {Cayrel} R., {Spite} M., {Spite} F.,
  {Chiappini} C., 2004, A\&A, 421, 613

\bibitem[{{Frenk} {et~al}\mbox{.}(1999){Frenk}, {White}, {Bode}, {Bond},
  {Bryan}, {Cen}, {Couchman}, {Evrard}, {Gnedin}, {Jenkins}, {Khokhlov},
  {Klypin}, {Navarro}, {Norman}, {Ostriker}, {Owen}, {Pearce}, {Pen},
  {Steinmetz}, {Thomas}, {Villumsen}, {Wadsley}, {Warren}, {Xu}, \&
  {Yepes}}]{frenk99}
{Frenk} C.~S. {et~al.}, 1999, ApJ, 525, 554

\bibitem[{{Gibson}(1997)}]{gibson97}
{Gibson} B.~K., 1997, MNRAS, 290, 471

\bibitem[{{Gibson}(1998)}]{gibson98}
{Gibson} B.~K., 1998, ApJ, 501, 675

\bibitem[{{Gibson} {et~al}\mbox{.}(2013){Gibson}, {Pilkington}, {Brook},
  {Stinson}, \& {Bailin}}]{gibson13}
{Gibson} B.~K., {Pilkington} K., {Brook} C.~B., {Stinson} G.~S., {Bailin} J.,
  2013, A\&A, 554, A47

\bibitem[{{Gratton} {et~al}\mbox{.}(2003){Gratton}, {Carretta}, {Claudi},
  {Lucatello}, \& {Barbieri}}]{gratton03}
{Gratton} R.~G., {Carretta} E., {Claudi} R., {Lucatello} S., {Barbieri} M.,
  2003, A\&A, 404, 187

\bibitem[{{Greggio} \& {Renzini}(1983)}]{greggio83}
{Greggio} L., {Renzini} A., 1983, A\&A, 118, 217

\bibitem[{{Guo} {et~al}\mbox{.}(2010){Guo}, {White}, {Li}, \&
  {Boylan-Kolchin}}]{guo10}
{Guo} Q., {White} S., {Li} C., {Boylan-Kolchin} M., 2010, MNRAS, 404, 1111

\bibitem[{{Haardt} \& {Madau}(1996)}]{haardt96}
{Haardt} F., {Madau} P., 1996, ApJ, 461, 20

\bibitem[{{Hachisu}, {Kato} \& {Nomoto}(1999){Hachisu}, {Kato}, \&
  {Nomoto}}]{hachisu99}
{Hachisu} I., {Kato} M., {Nomoto} K., 1999, ApJ, 522, 487

\bibitem[{{Haywood} {et~al}\mbox{.}(2013){Haywood}, {Di Matteo}, {Lehnert},
  {Katz}, \& {G{\'o}mez}}]{haywood13}
{Haywood} M., {Di Matteo} P., {Lehnert} M.~D., {Katz} D., {G{\'o}mez} A., 2013,
  A\&A, 560, A109

\bibitem[{{House} {et~al}\mbox{.}(2011){House}, {Brook}, {Gibson},
  {S{\'a}nchez-Bl{\'a}zquez}, {Courty}, {Few}, {Governato}, {Kawata}, {Ro{\v
  s}kar}, {Steinmetz}, {Stinson}, \& {Teyssier}}]{house11}
{House} E.~L. {et~al.}, 2011, MNRAS, 415, 2652

\bibitem[{{Iben} \& {Truran}(1978)}]{iben78}
{Iben}, Jr. I., {Truran} J.~W., 1978, ApJ, 220, 980

\bibitem[{{Iwamoto} {et~al}\mbox{.}(1999){Iwamoto}, {Brachwitz}, {Nomoto},
  {Kishimoto}, {Umeda}, {Hix}, \& {Thielemann}}]{iwamoto99}
{Iwamoto} K., {Brachwitz} F., {Nomoto} K., {Kishimoto} N., {Umeda} H., {Hix}
  W.~R., {Thielemann} F., 1999, ApJs, 125, 439

\bibitem[{{Izzard} {et~al}\mbox{.}(2004){Izzard}, {Tout}, {Karakas}, \&
  {Pols}}]{izzard04}
{Izzard} R.~G., {Tout} C.~A., {Karakas} A.~I., {Pols} O.~R., 2004, MNRAS, 350,
  407

\bibitem[{{Johnson}, {Ivans} \& {Stetson}(2006){Johnson}, {Ivans}, \&
  {Stetson}}]{johnson06}
{Johnson} J.~A., {Ivans} I.~I., {Stetson} P.~B., 2006, ApJ, 640, 801

\bibitem[{{Jones} {et~al}\mbox{.}(2013){Jones}, {Ellis}, {Richard}, \&
  {Jullo}}]{jones13}
{Jones} T., {Ellis} R.~S., {Richard} J., {Jullo} E., 2013, ApJ, 765, 48

\bibitem[{{Karakas} \& {Lattanzio}(2007)}]{karakas07}
{Karakas} A., {Lattanzio} J.~C., 2007, PASA, 24, 103

\bibitem[{{Karakas}(2010)}]{karakas10}
{Karakas} A.~I., 2010, MNRAS, 403, 1413

\bibitem[{{Kawata} \& {Gibson}(2003)}]{kawata03}
{Kawata} D., {Gibson} B.~K., 2003, MNRAS, 340, 908

\bibitem[{{Kennicutt}(1998)}]{kennicutt98}
{Kennicutt}, Jr. R.~C., 1998, ApJ, 498, 541

\bibitem[{{Knebe} {et~al}\mbox{.}(2000){Knebe}, {Kravtsov}, {Gottl{\"o}ber}, \&
  {Klypin}}]{knebe00}
{Knebe} A., {Kravtsov} A.~V., {Gottl{\"o}ber} S., {Klypin} A.~A., 2000, MNRAS,
  317, 630

\bibitem[{{Kobayashi}(2004)}]{kobayashi04}
{Kobayashi} C., 2004, MNRAS, 347, 740

\bibitem[{{Kobayashi} \& {Nakasato}(2011)}]{kobayashi11}
{Kobayashi} C., {Nakasato} N., 2011, ApJ, 729, 16

\bibitem[{{Kobayashi}, {Tsujimoto} \& {Nomoto}(2000){Kobayashi}, {Tsujimoto},
  \& {Nomoto}}]{kobayashi00}
{Kobayashi} C., {Tsujimoto} T., {Nomoto} K., 2000, ApJ, 539, 26

\bibitem[{{Kobayashi} {et~al}\mbox{.}(1998){Kobayashi}, {Tsujimoto}, {Nomoto},
  {Hachisu}, \& {Kato}}]{kobayashi98}
{Kobayashi} C., {Tsujimoto} T., {Nomoto} K., {Hachisu} I., {Kato} M., 1998,
  ApJl, 503, L155

\bibitem[{{Kobayashi} {et~al}\mbox{.}(2006){Kobayashi}, {Umeda}, {Nomoto},
  {Tominaga}, \& {Ohkubo}}]{kobayashi06}
{Kobayashi} C., {Umeda} H., {Nomoto} K., {Tominaga} N., {Ohkubo} T., 2006, ApJ,
  653, 1145

\bibitem[{{Kodama} \& {Arimoto}(1997)}]{kodama97}
{Kodama} T., {Arimoto} N., 1997, A\&A, 320, 41

\bibitem[{{Kroupa}(2001)}]{kroupa01}
{Kroupa} P., 2001, MNRAS, 322, 231

\bibitem[{{Kroupa}, {Tout} \& {Gilmore}(1993){Kroupa}, {Tout}, \&
  {Gilmore}}]{kroupa93}
{Kroupa} P., {Tout} C.~A., {Gilmore} G., 1993, MNRAS, 262, 545

\bibitem[{{Lapenna} {et~al}\mbox{.}(2012){Lapenna}, {Mucciarelli}, {Origlia},
  \& {Ferraro}}]{lapenna12}
{Lapenna} E., {Mucciarelli} A., {Origlia} L., {Ferraro} F.~R., 2012, ApJ, 761,
  33

\bibitem[{{Lia}, {Portinari} \& {Carraro}(2002){Lia}, {Portinari}, \&
  {Carraro}}]{lia02}
{Lia} C., {Portinari} L., {Carraro} G., 2002, MNRAS, 330, 821

\bibitem[{{Limongi} \& {Chieffi}(2003)}]{limongi03}
{Limongi} M., {Chieffi} A., 2003, ApJ, 592, 404

\bibitem[{{Maciel}, {Costa} \& {Uchida}(2003){Maciel}, {Costa}, \&
  {Uchida}}]{maciel03}
{Maciel} W.~J., {Costa} R.~D.~D., {Uchida} M.~M.~M., 2003, A\&A, 397, 667

\bibitem[{{Madau}, {della Valle} \& {Panagia}(1998){Madau}, {della Valle}, \&
  {Panagia}}]{madau98}
{Madau} P., {della Valle} M., {Panagia} N., 1998, MNRAS, 297, L17

\bibitem[{{Maeder}(1992)}]{maeder92}
{Maeder} A., 1992, A\&A, 264, 105

\bibitem[{{Mannucci}, {Della Valle} \& {Panagia}(2006){Mannucci}, {Della
  Valle}, \& {Panagia}}]{mannucci06}
{Mannucci} F., {Della Valle} M., {Panagia} N., 2006, MNRAS, 370, 773

\bibitem[{{Mannucci} {et~al}\mbox{.}(2005){Mannucci}, {Della Valle}, {Panagia},
  {Cappellaro}, {Cresci}, {Maiolino}, {Petrosian}, \& {Turatto}}]{mannucci05}
{Mannucci} F., {Della Valle} M., {Panagia} N., {Cappellaro} E., {Cresci} G.,
  {Maiolino} R., {Petrosian} A., {Turatto} M., 2005, A\&A, 433, 807

\bibitem[{{Mannucci} {et~al}\mbox{.}(2008){Mannucci}, {Maoz}, {Sharon},
  {Botticella}, {Della Valle}, {Gal-Yam}, \& {Panagia}}]{mannucci08}
{Mannucci} F., {Maoz} D., {Sharon} K., {Botticella} M.~T., {Della Valle} M.,
  {Gal-Yam} A., {Panagia} N., 2008, MNRAS, 383, 1121

\bibitem[{{Maoz}(2008)}]{maoz08}
{Maoz} D., 2008, MNRAS, 384, 267

\bibitem[{{Marigo}(2001)}]{marigo01}
{Marigo} P., 2001, A\&A, 370, 194

\bibitem[{{Martel}, {Kawata} \& {Ellison}(2013){Martel}, {Kawata}, \&
  {Ellison}}]{martel13}
{Martel} H., {Kawata} D., {Ellison} S.~L., 2013, MNRAS, 431, 2560

\bibitem[{{Mart{\'{\i}}nez-Serrano}
  {et~al}\mbox{.}(2008){Mart{\'{\i}}nez-Serrano}, {Serna},
  {Dom{\'{\i}}nguez-Tenreiro}, \& {Moll{\'a}}}]{martinezserrano08a}
{Mart{\'{\i}}nez-Serrano} F.~J., {Serna} A., {Dom{\'{\i}}nguez-Tenreiro} R.,
  {Moll{\'a}} M., 2008, MNRAS, 388, 39

\bibitem[{{Matteucci} \& {Francois}(1989)}]{matteucci89}
{Matteucci} F., {Francois} P., 1989, MNRAS, 239, 885

\bibitem[{{Matteucci} \& {Greggio}(1986)}]{matteucci86}
{Matteucci} F., {Greggio} L., 1986, A\&A, 154, 279

\bibitem[{{Matteucci} \& {Recchi}(2001)}]{matteucci01}
{Matteucci} F., {Recchi} S., 2001, ApJ, 558, 351

\bibitem[{{Matteucci} {et~al}\mbox{.}(2009){Matteucci}, {Spitoni}, {Recchi}, \&
  {Valiante}}]{matteucci09}
{Matteucci} F., {Spitoni} E., {Recchi} S., {Valiante} R., 2009, A\&A, 501, 531

\bibitem[{{Minchev}, {Chiappini} \& {Martig}(2013){Minchev}, {Chiappini}, \&
  {Martig}}]{minchev13}
{Minchev} I., {Chiappini} C., {Martig} M., 2013, A\&A, 558, A9

\bibitem[{{Nomoto} {et~al}\mbox{.}(1997){Nomoto}, {Hashimoto}, {Tsujimoto},
  {Thielemann}, {Kishimoto}, {Kubo}, \& {Nakasato}}]{nomoto97}
{Nomoto} K., {Hashimoto} M., {Tsujimoto} T., {Thielemann} F.-K., {Kishimoto}
  N., {Kubo} Y., {Nakasato} N., 1997, Nuclear Physics A, 616, 79

\bibitem[{{Nomoto}, {Thielemann} \& {Wheeler}(1984){Nomoto}, {Thielemann}, \&
  {Wheeler}}]{nomoto84}
{Nomoto} K., {Thielemann} F.-K., {Wheeler} J.~C., 1984, ApJl, 279, L23

\bibitem[{{Oppenheimer} \& {Dav{\'e}}(2008)}]{oppenheimer08}
{Oppenheimer} B.~D., {Dav{\'e}} R., 2008, MNRAS, 387, 577

\bibitem[{{O'Shea} {et~al}\mbox{.}(2005){O'Shea}, {Nagamine}, {Springel},
  {Hernquist}, \& {Norman}}]{oshea05}
{O'Shea} B.~W., {Nagamine} K., {Springel} V., {Hernquist} L., {Norman} M.~L.,
  2005, ApJs, 160, 1

\bibitem[{{Pagel} \& {Patchett}(1975)}]{pagel75}
{Pagel} B.~E.~J., {Patchett} B.~E., 1975, MNRAS, 172, 13

\bibitem[{{Pignatari} {et~al}\mbox{.}(2013){Pignatari}, {Herwig}, {Hirschi},
  {Bennett}, {Rockefeller}, {Fryer}, {Timmes}, {Heger}, {Jones}, {Battino},
  {Ritter}, {Dotter}, {Trappitsch}, {Diehl}, {Frischknecht}, {Hungerford},
  {Magkotsios}, {Travaglio}, \& {Young}}]{pignatari13}
{Pignatari} M. {et~al.}, 2013, arXiv: 1307.6961

\bibitem[{{Pilkington} {et~al}\mbox{.}(2012{\natexlab{a}}){Pilkington}, {Few},
  {Gibson}, {Calura}, {Michel-Dansac}, {Thacker}, {Moll{\'a}}, {Matteucci},
  {Rahimi}, {Kawata}, {Kobayashi}, {Brook}, {Stinson}, {Couchman}, {Bailin}, \&
  {Wadsley}}]{pilkington12a}
{Pilkington} K. {et~al.}, 2012{\natexlab{a}}, A\&A, 540, A56

\bibitem[{{Pilkington} {et~al}\mbox{.}(2012{\natexlab{b}}){Pilkington},
  {Gibson}, {Brook}, {Calura}, {Stinson}, {Thacker}, {Michel-Dansac}, {Bailin},
  {Couchman}, {Wadsley}, {Quinn}, \& {Maccio}}]{pilkington12c}
{Pilkington} K. {et~al.}, 2012{\natexlab{b}}, MNRAS, 425, 969

\bibitem[{{Portinari}, {Chiosi} \& {Bressan}(1998){Portinari}, {Chiosi}, \&
  {Bressan}}]{portinari98}
{Portinari} L., {Chiosi} C., {Bressan} A., 1998, A\&A, 334, 505

\bibitem[{{Prieto}, {Stanek} \& {Beacom}(2008){Prieto}, {Stanek}, \&
  {Beacom}}]{prieto08}
{Prieto} J.~L., {Stanek} K.~Z., {Beacom} J.~F., 2008, ApJ, 673, 999

\bibitem[{{Quimby} {et~al}\mbox{.}(2006){Quimby}, {H{\"o}flich}, {Kannappan},
  {Rykoff}, {Rujopakarn}, {Akerlof}, {Gerardy}, \& {Wheeler}}]{quimby06}
{Quimby} R., {H{\"o}flich} P., {Kannappan} S.~J., {Rykoff} E., {Rujopakarn} W.,
  {Akerlof} C.~W., {Gerardy} C.~L., {Wheeler} J.~C., 2006, ApJ, 636, 400

\bibitem[{{Ram{\'{\i}}rez}, {Allende Prieto} \&
  {Lambert}(2007){Ram{\'{\i}}rez}, {Allende Prieto}, \& {Lambert}}]{ramirez07}
{Ram{\'{\i}}rez} I., {Allende Prieto} C., {Lambert} D.~L., 2007, A\&A, 465, 271

\bibitem[{{Reddy}, {Lambert} \& {Allende Prieto}(2006){Reddy}, {Lambert}, \&
  {Allende Prieto}}]{reddy06}
{Reddy} B.~E., {Lambert} D.~L., {Allende Prieto} C., 2006, MNRAS, 367, 1329

\bibitem[{{Reddy} {et~al}\mbox{.}(2003){Reddy}, {Tomkin}, {Lambert}, \&
  {Allende Prieto}}]{reddy03}
{Reddy} B.~E., {Tomkin} J., {Lambert} D.~L., {Allende Prieto} C., 2003, MNRAS,
  340, 304

\bibitem[{{Romano} {et~al}\mbox{.}(2010){Romano}, {Karakas}, {Tosi}, \&
  {Matteucci}}]{romano10}
{Romano} D., {Karakas} A.~I., {Tosi} M., {Matteucci} F., 2010, A\&A, 522, A32

\bibitem[{{Romeo}, {Portinari} \& {Sommer-Larsen}(2005){Romeo}, {Portinari}, \&
  {Sommer-Larsen}}]{romeo05}
{Romeo} A.~D., {Portinari} L., {Sommer-Larsen} J., 2005, MNRAS, 361, 983

\bibitem[{{Rosen} \& {Bregman}(1995)}]{rosen95}
{Rosen} A., {Bregman} J.~N., 1995, ApJ, 440, 634

\bibitem[{{Salpeter}(1955)}]{salpeter55}
{Salpeter} E.~E., 1955, ApJ, 121, 161

\bibitem[{{S{\'a}nchez-Bl{\'a}zquez}
  {et~al}\mbox{.}(2009){S{\'a}nchez-Bl{\'a}zquez}, {Courty}, {Gibson}, \&
  {Brook}}]{sanchezblazquez09}
{S{\'a}nchez-Bl{\'a}zquez} P., {Courty} S., {Gibson} B.~K., {Brook} C.~B.,
  2009, MNRAS, 398, 591

\bibitem[{{Scalo}(1998)}]{scalo98}
{Scalo} J., 1998, in Astronomical Society of the Pacific Conference Series,
  Vol. 142, The Stellar Initial Mass Function (38th Herstmonceux Conference),
  {Gilmore} G., {Howell} D., eds., p. 201

\bibitem[{{Scalo}(1986)}]{scalo86}
{Scalo} J.~M., 1986, FCPh, 11, 1

\bibitem[{{Scannapieco} {et~al}\mbox{.}(2005){Scannapieco}, {Tissera}, {White},
  \& {Springel}}]{scannapieco05}
{Scannapieco} C., {Tissera} P.~B., {White} S.~D.~M., {Springel} V., 2005,
  MNRAS, 364, 552

\bibitem[{{Scannapieco} {et~al}\mbox{.}(2012){Scannapieco}, {Wadepuhl},
  {Parry}, {Navarro}, {Jenkins}, {Springel}, {Teyssier}, {Carlson}, {Couchman},
  {Crain}, {Dalla Vecchia}, {Frenk}, {Kobayashi}, {Monaco}, {Murante},
  {Okamoto}, {Quinn}, {Schaye}, {Stinson}, {Theuns}, {Wadsley}, {White}, \&
  {Woods}}]{scannapieco12}
{Scannapieco} C. {et~al.}, 2012, MNRAS, 423, 1726

\bibitem[{{Scannapieco} \& {Bildsten}(2005)}]{scannapieco05b}
{Scannapieco} E., {Bildsten} L., 2005, ApJl, 629, L85

\bibitem[{{Schmidt}(1959)}]{schmidt59}
{Schmidt} M., 1959, ApJ, 129, 243

\bibitem[{{Sch{\"o}nrich} \& {Binney}(2009)}]{schoenrich09}
{Sch{\"o}nrich} R., {Binney} J., 2009, MNRAS, 399, 1145

\bibitem[{{Sharon} {et~al}\mbox{.}(2007){Sharon}, {Gal-Yam}, {Maoz},
  {Filippenko}, \& {Guhathakurta}}]{sharon07}
{Sharon} K., {Gal-Yam} A., {Maoz} D., {Filippenko} A.~V., {Guhathakurta} P.,
  2007, ApJ, 660, 1165

\bibitem[{{Shen}, {Wadsley} \& {Stinson}(2010){Shen}, {Wadsley}, \&
  {Stinson}}]{shen10}
{Shen} S., {Wadsley} J., {Stinson} G., 2010, MNRAS, 407, 1581

\bibitem[{{Stinson} {et~al}\mbox{.}(2013){Stinson}, {Bovy}, {Rix}, {Brook},
  {Ro{\v s}kar}, {Dalcanton}, {Macci{\`o}}, {Wadsley}, {Couchman}, \&
  {Quinn}}]{stinson13}
{Stinson} G.~S. {et~al.}, 2013, MNRAS, 436, 625

\bibitem[{{Strolger} {et~al}\mbox{.}(2004){Strolger}, {Riess}, {Dahlen},
  {Livio}, {Panagia}, {Challis}, {Tonry}, {Filippenko}, {Chornock}, {Ferguson},
  {Koekemoer}, {Mobasher}, {Dickinson}, {Giavalisco}, {Casertano}, {Hook},
  {Blondin}, {Leibundgut}, {Nonino}, {Rosati}, {Spinrad}, {Steidel}, {Stern},
  {Garnavich}, {Matheson}, {Grogin}, {Hornschemeier}, {Kretchmer}, {Laidler},
  {Lee}, {Lucas}, {de Mello}, {Moustakas}, {Ravindranath}, {Richardson}, \&
  {Taylor}}]{strolger04}
{Strolger} L.-G. {et~al.}, 2004, ApJ, 613, 200

\bibitem[{{Strolger} {et~al}\mbox{.}(2002){Strolger}, {Smith}, {Suntzeff},
  {Phillips}, {Aldering}, {Nugent}, {Knop}, {Perlmutter}, {Schommer}, {Ho},
  {Hamuy}, {Krisciunas}, {Germany}, {Covarrubias}, {Candia}, {Athey}, {Blanc},
  {Bonacic}, {Bowers}, {Conley}, {Dahl{\'e}n}, {Freedman}, {Galaz}, {Gates},
  {Goldhaber}, {Goobar}, {Groom}, {Hook}, {Marzke}, {Mateo}, {McCarthy},
  {M{\'e}ndez}, {Muena}, {Persson}, {Quimby}, {Roth}, {Ruiz-Lapuente},
  {Seguel}, {Szentgyorgyi}, {von Braun}, {Wood-Vasey}, \& {York}}]{strolger02}
{Strolger} L.-G. {et~al.}, 2002, AJ, 124, 2905

\bibitem[{{Suda} {et~al}\mbox{.}(2013){Suda}, {Komiya}, {Yamada}, {Katsuta},
  {Aoki}, {Gil-Pons}, {Doherty}, {Campbell}, {Wood}, \& {Fujimoto}}]{suda13}
{Suda} T. {et~al.}, 2013, MNRAS, 432, L46

\bibitem[{{Sullivan} {et~al}\mbox{.}(2006){Sullivan}, {Le Borgne}, {Pritchet},
  {Hodsman}, {Neill}, {Howell}, {Carlberg}, {Astier}, {Aubourg}, {Balam},
  {Basa}, {Conley}, {Fabbro}, {Fouchez}, {Guy}, {Hook}, {Pain},
  {Palanque-Delabrouille}, {Perrett}, {Regnault}, {Rich}, {Taillet}, {Baumont},
  {Bronder}, {Ellis}, {Filiol}, {Lusset}, {Perlmutter}, {Ripoche}, \&
  {Tao}}]{sullivan06}
{Sullivan} M. {et~al.}, 2006, ApJ, 648, 868

\bibitem[{{Talbot} \& {Arnett}(1971)}]{talbot71}
{Talbot}, Jr. R.~J., {Arnett} W.~D., 1971, ApJ, 170, 409

\bibitem[{{Tasker} {et~al}\mbox{.}(2008){Tasker}, {Brunino}, {Mitchell},
  {Michielsen}, {Hopton}, {Pearce}, {Bryan}, \& {Theuns}}]{tasker08}
{Tasker} E.~J., {Brunino} R., {Mitchell} N.~L., {Michielsen} D., {Hopton} S.,
  {Pearce} F.~R., {Bryan} G.~L., {Theuns} T., 2008, MNRAS, 390, 1267

\bibitem[{{Teyssier}(2002)}]{teyssier02}
{Teyssier} R., 2002, A\&A, 385, 337

\bibitem[{{Teyssier} {et~al}\mbox{.}(2013){Teyssier}, {Pontzen}, {Dubois}, \&
  {Read}}]{teyssier13}
{Teyssier} R., {Pontzen} A., {Dubois} Y., {Read} J.~I., 2013, MNRAS, 429, 3068

\bibitem[{{Timmes}, {Woosley} \& {Weaver}(1995){Timmes}, {Woosley}, \&
  {Weaver}}]{timmes95}
{Timmes} F.~X., {Woosley} S.~E., {Weaver} T.~A., 1995, ApJs, 98, 617

\bibitem[{{Tinsley}(1980)}]{tinsley80}
{Tinsley} B.~M., 1980, FCPh, 5, 287

\bibitem[{{Tornatore} {et~al}\mbox{.}(2007){Tornatore}, {Borgani}, {Dolag}, \&
  {Matteucci}}]{tornatore07}
{Tornatore} L., {Borgani} S., {Dolag} K., {Matteucci} F., 2007, MNRAS, 382,
  1050

\bibitem[{{Tornatore} {et~al}\mbox{.}(2004){Tornatore}, {Borgani}, {Matteucci},
  {Recchi}, \& {Tozzi}}]{tornatore04}
{Tornatore} L., {Borgani} S., {Matteucci} F., {Recchi} S., {Tozzi} P., 2004,
  MNRAS, 349, L19

\bibitem[{{Tornatore} {et~al}\mbox{.}(2010){Tornatore}, {Borgani}, {Viel}, \&
  {Springel}}]{tornatore10}
{Tornatore} L., {Borgani} S., {Viel} M., {Springel} V., 2010, MNRAS, 402, 1911

\bibitem[{{Truelove} {et~al}\mbox{.}(1997){Truelove}, {Klein}, {McKee},
  {Holliman}, {Howell}, \& {Greenough}}]{truelove97}
{Truelove} J.~K., {Klein} R.~I., {McKee} C.~F., {Holliman}, II J.~H., {Howell}
  L.~H., {Greenough} J.~A., 1997, ApJl, 489, L179

\bibitem[{{Tsujimoto} {et~al}\mbox{.}(1995){Tsujimoto}, {Nomoto}, {Yoshii},
  {Hashimoto}, {Yanagida}, \& {Thielemann}}]{tsujimoto95}
{Tsujimoto} T., {Nomoto} K., {Yoshii} Y., {Hashimoto} M., {Yanagida} S.,
  {Thielemann} F.-K., 1995, MNRAS, 277, 945

\bibitem[{{Tutukov} \& {Yungelson}(1994)}]{tutukov94}
{Tutukov} A.~V., {Yungelson} L.~R., 1994, MNRAS, 268, 871

\bibitem[{{Valdarnini}(2003)}]{valdarnini03}
{Valdarnini} R., 2003, MNRAS, 339, 1117

\bibitem[{{van den Hoek} \& {Groenewegen}(1997)}]{vdhoek97}
{van den Hoek} L.~B., {Groenewegen} M.~A.~T., 1997, A\&As, 123, 305

\bibitem[{{Van der Swaelmen} {et~al}\mbox{.}(2013){Van der Swaelmen}, {Hill},
  {Primas}, \& {Cole}}]{vdswaelmen13}
{Van der Swaelmen} M., {Hill} V., {Primas} F., {Cole} A.~A., 2013, A\&A, 560,
  A44

\bibitem[{{Wiersma} {et~al}\mbox{.}(2009){Wiersma}, {Schaye}, {Theuns}, {Dalla
  Vecchia}, \& {Tornatore}}]{wiersma09}
{Wiersma} R.~P.~C., {Schaye} J., {Theuns} T., {Dalla Vecchia} C., {Tornatore}
  L., 2009, MNRAS, 399, 574

\bibitem[{{Woosley} \& {Weaver}(1995)}]{ww95}
{Woosley} S.~E., {Weaver} T.~A., 1995, ApJs, 101, 181

\bibitem[{{Yates} {et~al}\mbox{.}(2013){Yates}, {Henriques}, {Thomas},
  {Kauffmann}, {Johansson}, \& {White}}]{yates13}
{Yates} R.~M., {Henriques} B., {Thomas} P.~A., {Kauffmann} G., {Johansson} J.,
  {White} S.~D.~M., 2013, MNRAS

\bibitem[{{Yuan} {et~al}\mbox{.}(2011){Yuan}, {Kewley}, {Swinbank}, {Richard},
  \& {Livermore}}]{yuan11}
{Yuan} T.-T., {Kewley} L.~J., {Swinbank} A.~M., {Richard} J., {Livermore}
  R.~C., 2011, ApJl, 732, L14

\bibitem[{{Yungelson} \& {Livio}(1998)}]{yungelson98}
{Yungelson} L., {Livio} M., 1998, ApJ, 497, 168

\bibitem[{{Zaritsky}, {Kennicutt} \& {Huchra}(1994){Zaritsky}, {Kennicutt}, \&
  {Huchra}}]{zaritsky94}
{Zaritsky} D., {Kennicutt}, Jr. R.~C., {Huchra} J.~P., 1994, ApJ, 420, 87

\end{thebibliography}

\appendix
\section{Chemical evolution model inputs}
\label{extrapolate}

\begin{figure}
\includegraphics[width=84mm]{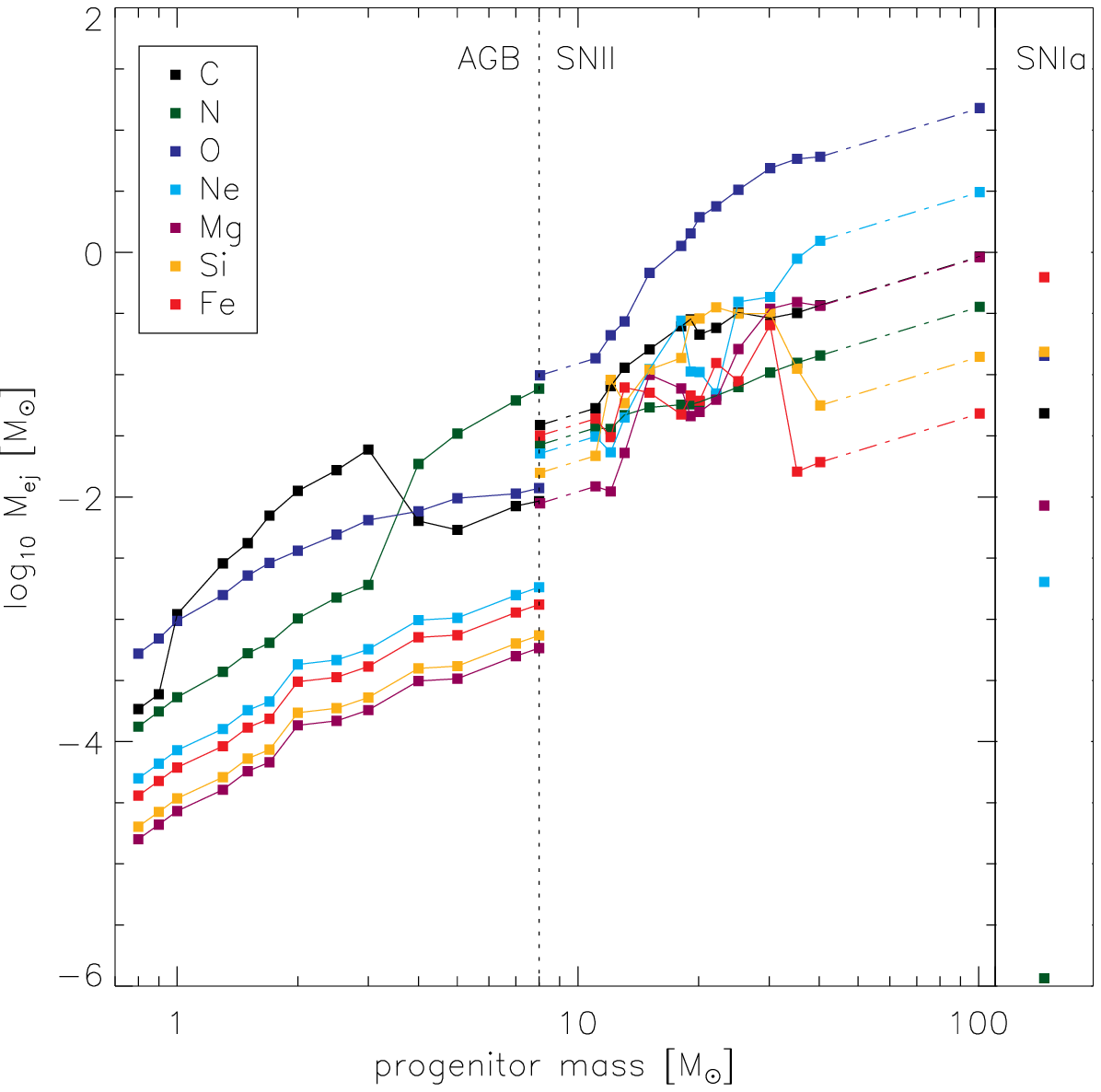}
\caption{Mass of elements ejected by stars as a function of initial
  mass. In the right-hand panel we show the abundances for a single
  SNIa for comparison, no labels are shown on the lower axis for SNIa
  as the ejecta elements are the same for all SNIa. The mass above
  which stars are considered to be SNeII progenitors is indicated at
  8~M$_\odot$ with a dotted line. Data for AGB stars are taken from \protect\cite{vdhoek97}, SNeII
 from \protect\cite{ww95}, and SNIa from \protect\cite{iwamoto99}. Points
  connected by solid lines denote the original data, those connected
  by dot-dash lines show adopted. Extrapolations are linear and scaled to the mass of the
  progenitor star.}
\label{fig:ejecta}
\end{figure}

In this work we employ the nucleosynthesis models of \cite{vdhoek97}
for AGB stars, \cite{ww95} for SNII and \cite{iwamoto99} for SNIa. 
The mass of different elements ejected from these sources is shown in
Fig.~\ref{fig:ejecta} as a function of the progenitor mass for stars
with an initially solar metallicity. The AGB yield from
\cite{vdhoek97} span a range of progenitor mass of 0.8--8~Msol, while
model grid from \cite{ww95} extends from a maximum of 40~M$_\odot$
down to 12~M$_\odot$ for most metallicities. To make use of these
nucleosynthesis models we interpolate between neighbouring grid points
within those models. A void exists between the mass ranges covered 
by the respective models. When necessary we extrapolate in mass by retaining the fractional
abundance of each element and scaling with the mass of the progenitor.
These extrapolations are illustrated in Fig.~\ref{fig:ejecta} with
dot-dashed lines.

The results of this work are clearly dependent on how we handle the
yields in the uncertain mass regimes, particularly regarding the most
massive stars. In the current version of our model we neglect fallback
in massive stars and hypernovae instead simply retaining the same
abundance ratios as 40~M$_\odot$ stars have. We note that this
extrapolation has no impact on our \emph{uM40} models for which all 
feedback is prevented for stars that are more massive. 

In our model we assume stars more massive than 8~M$_\odot$ explode as
SNII, the yields of the least massive SNII are obtain by assuming the
same yield as the least massive SNII in \cite{ww95} but scaling down
with the progenitors mass. In truth this is probably an over
estimation of the yields of these stars but this mass range
represents only 1.4--2.5\% of the mass of stars formed so it should not
have a large impact.

No extrapolation is required to lower masses as the least massive
star that ejects mass within the Hubble time is greater than
0.8~M$_\odot$ and thus within the mass bounds of the \cite{vdhoek97} models.

\section{Supplementary galaxies}

We performed several additional simulations that are not required in
the main text to explain the trends observed. To present our results it
was only necessary to show how the SNIa model \emph{IaK3} affected 
a single IMF, for which we chose \emph{S55}. We now also
demonstrate the effect of enhanced SNIa feedback on the \emph{K01} IMF.
First we show the \emph{K01-uM100-IaK} model from our main results
again alongside a realisation where we changed the SNIa model to the
enhanced version (\emph{IaK3}), we label this new realisation \emph{K01-uM100-IaK3}.
A second model is the same in most respects but is subject to a
reduced SNII upper mass limit of $m_\mathrm{SNII,u}$=80~M$_\odot$
which is labelled as \emph{K01-uM80-IaK3}. 

\begin{figure*}
\includegraphics[width=1.0\textwidth]{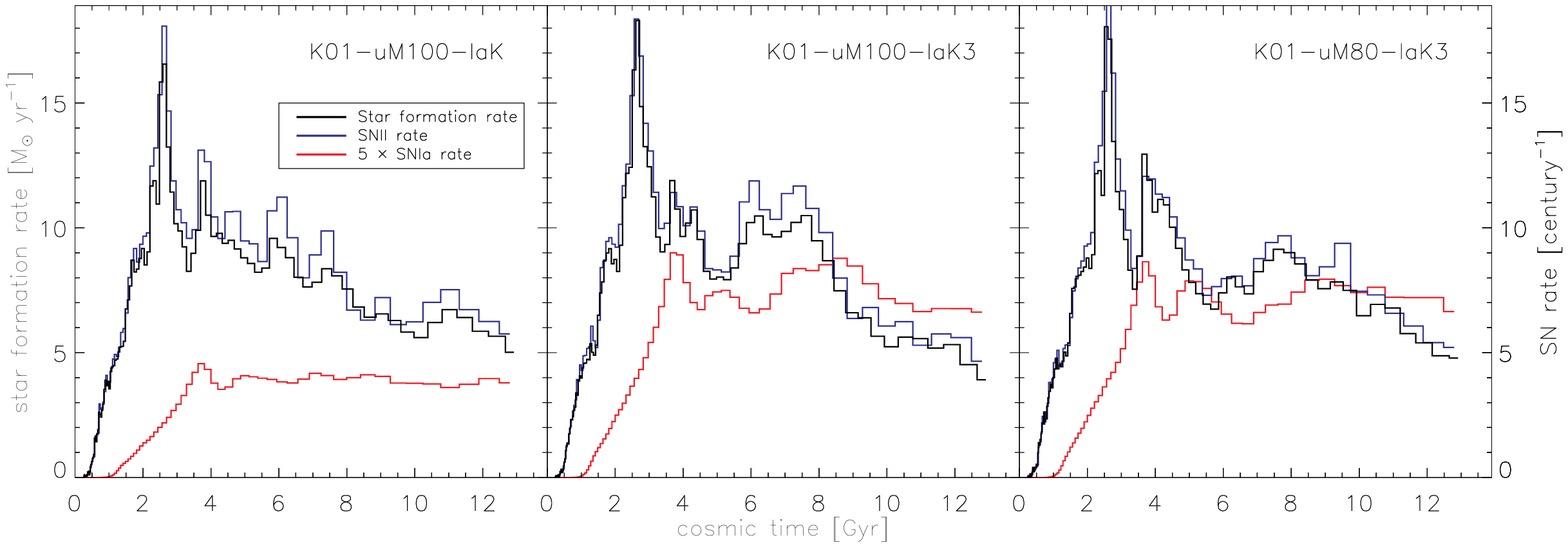}
\caption{The same as Fig.~\ref{fig:sncomp} but for three realisations not 
  discussed in the main text. The black
  line indicates the star formation rate with the scale given on the
  left axes. The SN rates are plotted in blue for SNII and red for
  SNIa. The units are given on the right axes but note that for
  clarity the SNIa rates are scaled up by a factor of 5.}
\label{fig:sncomp2}
\end{figure*}

The star formation and supernovae rates of these three realisations
are shown in Fig.~\ref{fig:sncomp2}. As is the case with the
\emph{S55-uM100-IaK3}, \emph{K01-uM100-IaK3} has a much higher SNIa
rate than its \emph{IaK} counterpart and the top-heavy nature of
the \emph{K01} IMF makes the impact even stronger which may be
responsible for the reduced star formation rate after 8~Gyr. The
second new model (\emph{k01-uM80-IaK3}) occupies the right-hand panel in
Fig.~\ref{fig:sncomp2} and has a similar SNIa rate as the previous model, in this case
the reduction in both SNII and SNIa feedback resulting from
the lower $m_\mathrm{SNII,u}$ allows the star formation rate to remain
slightly higher.

\begin{figure*}
\includegraphics[width=1.0\textwidth]{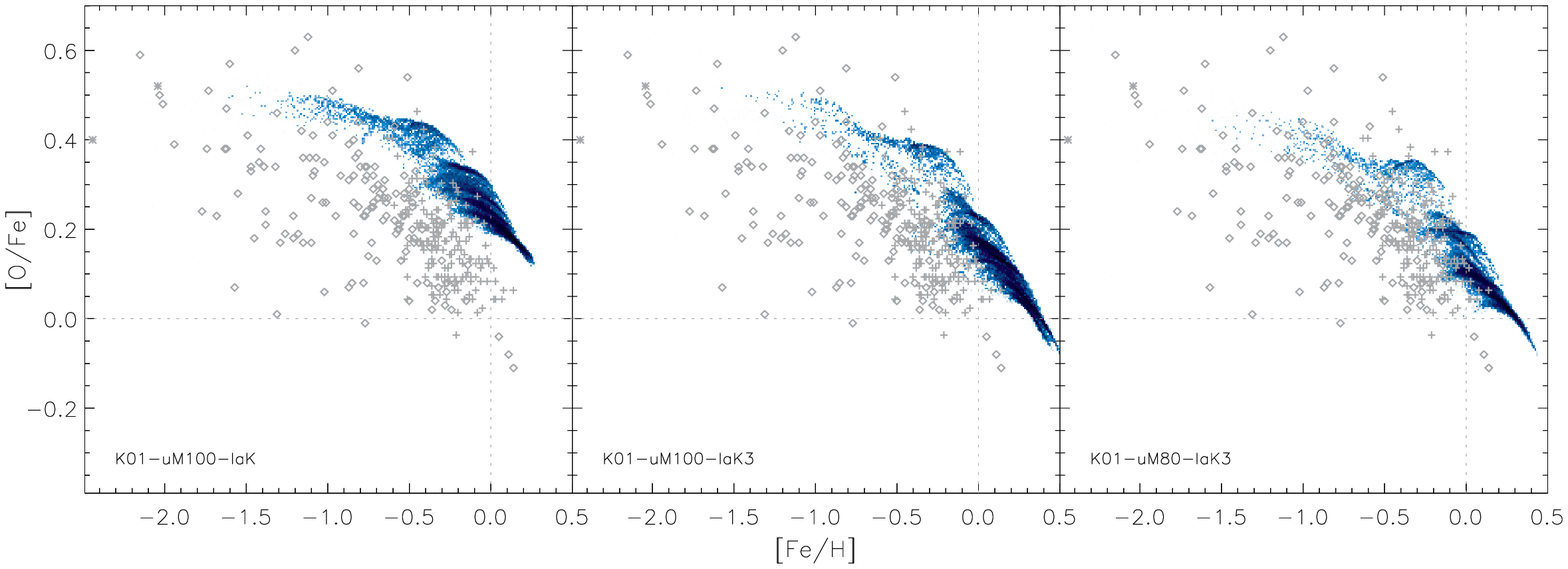}
\caption{The same as Fig.~\ref{fig:ofe} but for three realisations not discussed 
  in the main text. Darker colours indicate higher relative frequency density. Dotted lines indicate solar abundances.
  Observations are plotted in gray, diamonds: thick disc and halo stars \citep{gratton03}, plus
  signs: F and G dwarf stars in the thick and thin disc \citep{reddy03},
  and asterisks: very metal-poor stars \citep{cayrel04}.}
\label{fig:ofe2}
\end{figure*}

We also show how the abundance ratio plane (shown in Fig.~\ref{fig:ofe2}) is affected by the
introduction of enhanced SNIa chemical feedback. The \emph{K01} models is affected in
much the same way as the \emph{S55} models are: the increased SNIa feedback
steepens the slope in the high-metallicity regime. When looking at the
SN rates in Fig.~\ref{fig:sncomp2} we noted that since the \emph{K01} IMF
is more top-heavy than \emph{S55} the impact is stronger. This is
again apparent in Fig.~\ref{fig:ofe2} as the Fe content of the galaxy
is much higher. While the [O/Fe] ratio is nicely reproduced in this
way the [Fe/H] values are much too high, something which is not seen
in any of our other models. The excessive production of Fe is somewhat
ameliorated in \emph{K01-uM80-IaK3} since both SNII and SNIa rates of
stellar particles are reduced.

\label{lastpage}
\end{document}